\documentclass[fleqn,usenatbib]{mnras}

\usepackage{newtxtext,newtxmath}
\usepackage{subfigure}

\usepackage[T1]{fontenc}

\DeclareRobustCommand{\VAN}[3]{#2}
\let\VANthebibliography\thebibliography
\def\thebibliography{\DeclareRobustCommand{\VAN}[3]{##3}\VANthebibliography}

\usepackage{graphicx}	
\usepackage{amsmath}	

\newcommand{\teff}{$T_\mathrm{eff}$}
\newcommand{\logg}{$\log g$}

\DeclareMathOperator\erf{erf}

\title[Tracing the Origin of N-rich Stars]{On the connection between nitrogen-enhanced field stars and the Galactic globular clusters}

\author[S. G. Kane et al.]{
Sarah G. Kane,$^{1}$\thanks{E-mail: sgk27@cam.ac.uk (SGK)}
Vasily Belokurov,$^{1}$,
Stephanie Monty$^{1}$, Holger Baumgardt$^{2}$, Carrie Filion$^{3}$, \newauthor Andrey Kravtsov$^{4}$, GyuChul Myeong$^{1}$, HanYuan Zhang$^{1}$, Elana Kane$^{1}$
\\
$^{1}$ Institute of Astronomy, University of Cambridge, Madingley Road, Cambridge CB3 0HA, UK\\
$^{2}$ School of Mathematics and Physics, The University of Queensland, St. Lucia, QLD 4072, Australia\\
$^{3}$ Center for Computational Astrophysics, Flatiron Institute, 162 Fifth Avenue, New York, NY 10010, USA\\
$^{4}$ Department of Astronomy and Astrophysics, The University of Chicago, Chicago, IL 60637, USA
}

\date{Accepted XXX. Received YYY; in original form ZZZ}

\pubyear{\the\year{}}

\usepackage{bm}

\begin{document}
\label{firstpage}
\pagerange{\pageref{firstpage}--\pageref{lastpage}}
\maketitle

\begin{abstract}
As sites of some of the most efficient star formation in the Universe, globular clusters (GCs) have long been hypothesized to be the building blocks of young galaxies. Within the Milky Way, our best tracers of the contribution of GCs to the proto-Galaxy are stars with such anomalous overabundance in nitrogen and depletion in oxygen ("high-[N/O] stars") that they can be identified as having originated in a cluster long after they have escaped. We identify associations between these high-[N/O] field stars and GCs using integrals of motion and metallicities and compare to chemically typical halo stars to quantify any excess association, enabling a population-level exploration of the formation sites of the nitrogen-enhanced stars in the field. Relative to the halo as a whole, high-[N/O] stars show stronger associations with the most initially massive, inner Galaxy GCs, suggesting that many nitrogen-rich stars formed in these environments. However, when compared to a sample matched in orbital energy, the excess largely disappears: high-[N/O] stars are, on average, no more associated with surviving GCs than energy-matched halo stars, despite their [N/O] abundances indicating GC origins, consistent with a scenario in which a substantial fraction of low-energy inner-halo stars originate in GCs, so an energy-matched control dilutes any differential excess. We argue that associations between high-[N/O] stars and their parent GCs are further weakened because dynamical friction and the Galactic bar have altered integrals of motion, limiting the reliability of precise present-day associations and, especially, individual star-to-cluster tagging.
\end{abstract}




\section{Introduction}
\label{sec:intro}

Recent observations have begun to offer tantalizing clues that massive, bound star clusters may have played a more central role in early galaxy formation than previously thought. Compact, few-parsec clumps with stellar masses of $10^{5\text{–}7}~M_\odot$ are now commonly resolved in galaxies at $z\sim6$–10, consistent with the expected properties of young globular-cluster analogues \citep{Adamo_2024,mowla_firefly_sparkle,Fujimoto2024}. Spectroscopic studies of similarly high-redshift systems reveal super-solar N/O ratios and light-element abundance patterns reminiscent of second-population globular-cluster stars—possible signatures of early self-enrichment within dense proto-clusters \citep{Ji_2025,Naidu_2025,Cameron_nitrogen_gnz11,Napolitano2024}. Closer to home, Galactic-archaeology surveys have uncovered hundreds of nitrogen-rich halo giants with matching chemical signatures \citep{Schiavon_2017,Horta_nrich_stars}, and chemo-dynamical modelling suggests that a significant fraction of the Milky Way’s pre-disk stellar mass may have originated in massive clusters that were subsequently stripped or destroyed \citep[][]{belokurov_kravstov_nitrogen}. Together, these population-level hints motivate a systematic reassessment of how much early star formation occurred in massive clusters. We could begin to connect the dots between high-redshift observations and present-day stellar populations, if we had a well-calibrated model of globular cluster formation and disruption embedded within the broader context of galaxy formation.

Yet even the best current simulations still handle the disruption phase schematically. Cosmological zoom-in suites such as E-MOSAICS \citep[][]{Pfeffer2018,Kruijssen2019} and the FIRE-based cluster extensions \citep[][]{Ma2020,Grudic2023,Rodriguez2023} track GCs inside live galaxy potentials, yet still compute tidal mass-loss with analytic “tidal-tensor’’ recipes whose free parameters can be tuned post-factum.  Higher-resolution cloud-scale zooms, including the GRIFFIN simulations \citep[][]{Lahen2020,Lahen2025}, resolve cluster birth inside GMCs at sub-pc resolution but integrate for only $\lesssim$200 Myr, so they do not follow long-term evaporation or shocking beyond the starburst phase.  Direct $N$-body runs \citep[e.g.][]{Baumgardt_Makino_2003,Wang2016} capture internal relaxation and external tides from first principles, but must assume an over-simplified, static Galactic potential and an imposed Galactic orbit instead of a self-consistent cosmological setting.  Monte-Carlo/analytic population models  \citep[e.g.][]{Li2018,Choksi2018,Chen2024} evolve thousands of clusters through calibrated disruption channels, yet their mass-loss coefficients can always be re-tuned to match $z=0$ data.  Complementary semi-analytic frameworks such as EMACSS \citep[][]{Alexander2012} model coupled internal–external evolution for entire GC systems, but likewise rely on adjustable prescriptions for shocks and tides \citep[][]{Gieles2008,Gieles2016}.  Disruption remains the least constrained aspect of every framework,  so we need external tests, such as linking the nitrogen-rich GC-escapee stars now roaming the halo back to their original clusters, to pin it down.

Chemo-dynamical tagging provides exactly that check, combining the few elemental “fingerprints” that survive cluster dissolution with each star’s present-day orbit to trace it back to a common origin. The idea of a chemical tag rests on the premise that stars lock in the detailed abundance pattern of their birth cloud, so matching those patterns can reveal common origins \citep[][]{Freeman2002}.  In practice, however, only a handful of elements provide real discriminating power. It has been recently shown that the high-dimensional “chemical space’’ effectively collapses to fewer than ten independent axes, with $\alpha$–, iron-peak and a few odd-Z elements carrying most of the information; many other species are either too noisy or too tightly correlated to be useful \citep[][]{Ness2018,Ness2019,Ness2022,Manea2024}.  Moreover, until recently, uniform high-precision measurements of even these key tracers were available only for small, heterogeneous samples.  Wide-area surveys such as APOGEE, GALAH and \emph{Gaia} BP/RP have changed the game by delivering millions of stars on a common scale, yet most of those spectra still yield just a subset of the informative elements with uncertainties of $\gtrsim0.1$ dex—too coarse for chemistry alone to pin down individual birth sites.  The way forward is to fold in dynamics: if a star that is chemically distinctive also shares metallicity and integrals of motion with a particular cluster, the combined evidence sharply reduces false matches.  This “chemo-dynamical’’ tagging, enabled by \emph{Gaia} astrometry and Galactic-potential modeling, promises to turn sparse chemical labels into robust birth associations and provides a new empirical benchmark for testing globular-cluster disruption physics.

Chemical tagging is especially successful when stars are chemically very unique. Fortunately, this is the case for many stars that originate in GCs. Once thought to be relatively simple examples of a single burst of highly efficient star formation, members of typical GCs exhibit a relatively small spread in metallicities \citep{Gratton_2004}, although precisely how small these variations are, particularly among the first generation, is an area of active research \citep{Marino_2019,Marino_2023,Legnardi_2022,Carretta_1G_metallicity,Carretta_1G_metallicity_2}. However, contrary to those original, simple theories, clusters above a certain initial mass \citep{Milone_2020, Lagioia_2024} are now known to exhibit a distinctive and universal light element anti-correlation indicative of the presence of multiple populations. Within this anti-correlation, the first population (1P) members of a cluster are chemically typical and resemble field stars from the GC host galaxy that formed at a similar metallicity. However, unlike their 1P counterparts, 2P stars within a cluster are unusually enhanced in nitrogen, aluminum, and sodium and depleted in oxygen, carbon, and often magnesium \citep{Gratton_2004,Carretta_2009b,Carretta_2009a,Carretta2010,Gratton_what_is_GC,milone_multiple_pops_GCs}--a pattern of light element abundances that is shared across GCs but not in the field or other structures. 

Thus, although the chemistry of 1P stars does not allow us to identify them as having originated in a GC if they have escaped, the distinctive abundances of 2P stars do allow for effective chemical tagging. The nitrogen overabundance of 2P stars is so unusual that this element in particular has been regularly used to identify stars in the Galactic field that once originated in the second generation of a cluster \citep{Martell_2011,Schiavon_2017, Horta_nrich_stars, belokurov_kravstov_nitrogen}. However, using multiple abundances associated with the light element anti-correlation allows for the most accurate tagging of GC-origin stars \citep[e.g., because additional selection criteria remove stars that are N-enhanced from other processes, see][]{belokurov_kravstov_nitrogen}.


With GCs speculated to have contributed significantly to early Galactic star formation, the MW halo should be a graveyard of stars from totally dissolved GCs and "escapees" from presently disrupting clusters, particularly towards the Galactic center. Although high nitrogen abundances, among other elements, provide a reliable indicator that a star formed in \emph{a} GC, these light element anomalies do not necessarily offer distinguishing power to identify \emph{which} cluster that was. Several works have thus already connected field stars with GCs of potential origin. \citet{Savino_Posti_2019} compared the actions and metallicities of GCs and CN-strong giants from \citet{Martell_2010, Martell_2011} and found at least 7 stars chemodynamically consistent with having originated in surviving clusters (escapees) and note that about half of their stars have no such strong associations, indicating that they may have formed in totally tidally disrupted GCs. They also note the presence of a population of the CN-strong, presumably GC-born stars on surprisingly disk-like orbits.
Following from this work, \citet{Hanke2020_method} and \citet{Xu2024_method} also used orbital actions and metallicities to tag back nitrogen-enhanced stars to their birth clusters.
Recently, \citet{Souza2024_terzan5_runaway} examined a star with 2P-type chemistry, particularly nitrogen- and aluminum-enhancement, in the inner Galaxy and identified it as having likely been stripped from the GC Terzan 5 by the Galactic bar based on its orbital properties.

Motivated by the goal of better understanding the co-evolution of GCs and the Galaxy, we revisit the connection between N-rich field stars and existing MW GCs using APOGEE and our recently published data set of N-rich stars and information from both chemistry and kinematics. Given the importance of GCs to star formation in the early Milky Way--and perhaps the early stages of galaxy formation more generally--better understanding the relationship between N-rich field stars and GCs provides unique insight into the evolution of both GCs and the Galaxy. While previous works have chemo-dynamically linked N-rich field stars to individual plausible (surviving) birth cluster(s), we aim to investigate at the population level the association between these field stars and the surviving GCs. To this end, we quantify the associations between N-rich field stars and GCs based upon their integrals of motion and metallicities. These associations are compared to those between clusters and chemically typical stars in the halo to attempt to identify which clusters contributed most significantly to the population of 2P stars in the halo--and, perhaps correlatively, to the early stellar mass buildup in the Galaxy overall. Our datasets, comprised of chemically typical and high-[N/O] field stars, are collected from APOGEE \citep[][Section~\ref{subsec:APOGEE_data}]{APOGEE_survey,APOGEE_DR17} and machine learning-based abundance derivations from \textit{Gaia} BP/RP spectra \citep[][Section~\ref{subsec:candidate-data}]{Kane_2024}. We associate GCs and field stars with three separate, but similar, methods that transform the errors on the objects' observables into probability density functions in the space of their integrals of motion and metallicities. In particular, we make these associations based upon combinations of the stars' and GCs' energies, the $z$-component of their angular momenta, their actions, and their metallicities, as is described in Sections~\ref{subsec:2D_method}~and~\ref{subsec:3D4D_method}. The results of these approaches, which largely agree with each other, are discussed in Section~\ref{sec:results}. Further discussions of the implications of our findings as well as important caveats in our method are included in Section~\ref{sec:discussion}. Finally, we summarize our conclusions in Section~\ref{sec:conclusions}.

\section{Data and Methods}
\label{sec:data_and_methods}

We use two separate datasets to identify high-[N/O] field giants with chemistry consistent with having originated in the second generation of GCs: abundances from APOGEE spectroscopy and inferred abundances from \textit{Gaia} BP/RP low resolution spectrophotometry developed in \citet{Kane_2024}. We detail these two datasets and the criteria we use to classify stars with 2P-type chemistry in Sections \ref{subsec:APOGEE_data} and \ref{subsec:candidate-data}, respectively. We then outline the algorithm we use to associate 2P-type stars in the field with GCs via their orbital properties in Sections~\ref{subsec:2D_method} and \ref{subsec:3D4D_method}.

\subsection{Data}
\label{subsec:data}

\subsubsection{APOGEE Field Giants}
\label{subsec:APOGEE_data}

Our first sample of halo giants is selected from APOGEE Data Release 17 \citep{APOGEE_survey,APOGEE_DR17} using the ASPCAP pipeline's stellar parameters and abundances \citep{ASPCAP_APOGEE}. To select data with reliable stellar parameters, we require that the flags $\texttt{STAR\_BAD}<0$, $\texttt{TEFF\_BAD}<0$, $\texttt{LOGG\_BAD}<0$ and $\textrm{SNR}>20$. We also require the APOGEE errors to be lower than 0.2 for the reported abundances of [Fe/H], [N/Fe], [O/Fe], and [Al/Fe]. 
Giants are selected with $\log g<3$ and $T_\textrm{eff}<5200$~K. The sample is restricted to nearby giants with APOGEE \texttt{astroNN}\footnote{\url{https://github.com/henrysky/astroNN}} distances \citep{astroNN-Leung_Bovy_2019} $<10$~kpc with $3\times\mathrm{error}<\mathrm{distance}$. Using \texttt{astroNN} distances, APOGEE radial velocities, and astrometry from \textit{Gaia} Data Release 3 \citep{gaia_mission,gaia_dr3}, we calculate stellar positions and velocities in Astropy's \citep{astropy:2013,astropy:2018,astropy:2022} default Galactocentric coordinates. Disk stars are excluded with a velocity cut of $v_\phi<160~\mathrm{km/s}$. The selection is further limited to stars with $\textrm{[Fe/H]}<-1.0$, consistent with the approximate metallicity regime prior to spin-up, the appearance of the Galactic disc \citep{belokurov_aurora, chandra_three_phase, Zhang_2024_disk, conroy_galactic_disk_h3} to probe the expected peak of clustered star formation \citep{belokurov_kravstov_nitrogen,Kane_2024}. Because we seek to associate field stars with surviving clusters, known GC members from the \citet{vasiliev_gc} catalog are excluded.

The high-[N/O], high-[Al/Fe] sample is then selected via cuts $\textrm{[N/O]}-\textrm{[N/O] error}>0.55$ and $\textrm{[Al/Fe]}>-0.1$ \citep[consistent with the observed chemistry of 2P members in GCs, distinct from the Galactic field, as noted in][]{belokurov_kravstov_nitrogen}. The [N/O] enrichment of these field stars is so high that evolutionary effects (e.g., from dredge up along the giant branch) are insufficient to explain the abundances, and formation in GCs has conventionally been assumed to be the origin. The result of these cuts yields 117 field giants in the APOGEE data with chemistry consistent with having originated in the second generation of a GC. The median stellar parameters of these high-[N/O] stars are $T_\mathrm{eff}=4738~\textrm{K}$, $\log g =1.55$, and $\textrm{[Fe/H]}=-1.30$. The chemically typical ''background sample," which we use as comparison and validation data (see Sec.~\ref{subsec:2D_method}) has the same cuts as the high-[N/O] sample with the exception of the [N/O] selection, which is modified to $\textrm{[N/O]}+\textrm{[N/O] error}<0.55$, and the [Al/Fe] cut, which is omitted altogether, to exclude 2P-type stars in the Galactic field. The background sample consists of 4822 field giants.

\subsubsection{High-[N/O] Candidates from \textit{Gaia} BP/RP Spectra}
\label{subsec:candidate-data}

In addition to using high-[N/O], high-[Al/Fe] stars identifiable via APOGEE, we also use a dataset of 2P-type candidates in the field constructed from stellar parameters and abundances inferred from \textit{Gaia} BP/RP spectroscopy in \citet{Kane_2024}. \textit{Gaia} BP/RP spectra are very low resolution ($R\approx30-100$) spectra available for over 200 million stars in Data Release 3 \citep{gaia_dr3, gaia_bprp_processing, Gaia_bprp_calibration}, and given the multitude of spectra available, there have been a plethora of studies deriving stellar parameters and abundances from them, particularly using data-driven or machine learning-based approaches \citep[see, for example,][]{andrae_metallicities,li_alpha_bprp,lucey_cemp,fallows_sanders_bprp,Hattori_2024,ArdernArentsen_2024,Yao2024_bprp, Khalatyan2024_bprp}. 

\begin{figure*}
	\includegraphics[width=7in, alt={Each probability density function (PDF) of a star appears as a distinct shape in the E-Lz plane, with the GC PDFs looking like clouds, the APOGEE N-rich star PDFs looking like streaks, and the BP/RP N-ruch star PDFs appearing as slightly more condensed streaks (thus with somewhat more constrained PDFs). The PDFs often overlap with each other, especially at low energies, where the Lz values are constrained to a narrow range. The GC PDFs span the whole range of angular momenta and orbital energies, from very bound clusters in the inner Galaxy to clusters with energy well above Solar. Most, though not all, of the PDFs of the high-[N/O] field stars from both APOGEE and the BP/RP spectra fall below the in situ/accreted boundary. The boundary is a non-monotonic function of Lz that begins at approximately Solar energy at low values of Lz, varies quadratically at Lz values centered at 0, and then increases quadratically at positive (prograde) values of Lz, including energies well above Solar. The BP/RP stars do not extend to energies below approximately -25000 km^2/s^2, although the APOGEE stars do.}]{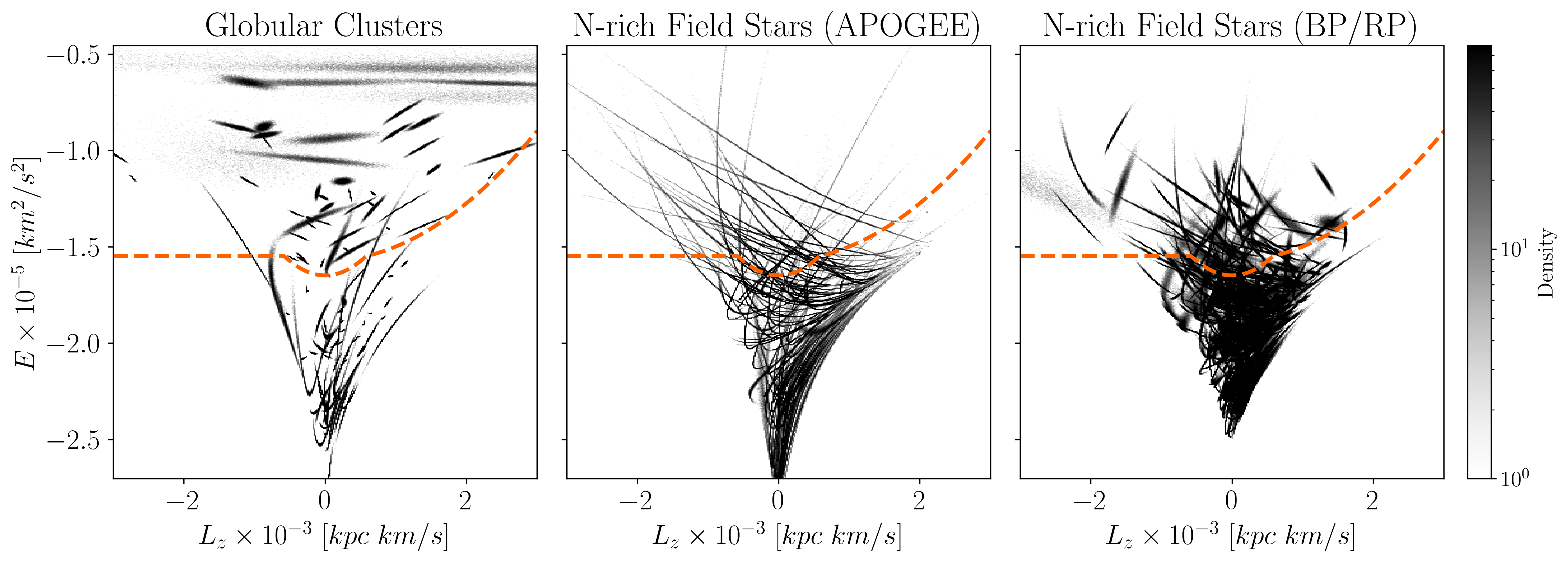}
    \caption{The 2D histograms of the distributions of the draws from the errors of the GCs (left), the high-[N/O] field stars from APOGEE (middle), and the high-[N/O] stars from the BP/RP sample (right) in $E-L_z$ space.  The individual ``streaks" or ``clouds" each represent a single object and are constructed from the errors on its observables. The red dashed line in each panel indicates the \textit{in-situ}/accreted boundary in this space from \citet{belokurov_kravstov_nitrogen} \citep[see also][]{Belokurov_2024_GCs}. A small number of objects are excluded by the $L_z$ range used for the visualization. Note that the GCs span a wider range of energies than either the APOGEE or BP/RP high-[N/O] stars, which are concentrated at lower energies.}
    \label{fig:ELz_hist}
\end{figure*}

Using a neural network, \citet{Kane_2024} perform heteroscedastic regression ---regression which takes into account non-uniform variances in the data --- to infer \teff, \logg, [Fe/H], [N/O], and [Al/Fe] and associated variances on those predictions from the BP/RP spectra. For consistency with that work, we refer to their network-predicted stellar parameters or abundances as $\mu_\textrm{value}$ (e.g., $\mu_{T_\mathrm{eff}}$, $\mu_{\textrm{[N/O]}}$, etc.) and the associated inferred standard deviation on that prediction as $\sigma_\textrm{value}$ ($\sigma_{T_\mathrm{eff}}$, $\sigma_{\textrm{[N/O]}}$, etc.). Because the neural network was trained with APOGEE labels, the abundances from the BP/RP spectra should be calibrated to the same scale; however, the BP/RP dataset has the benefit of being much larger than that from APOGEE with over six times more high-[N/O] field stars.

We apply the same velocity cut at $v_\phi<160~\mathrm{km}/\mathrm{s}$ and metallicity cut at $\textrm{[Fe/H]}<-1$ as was used for the APOGEE data to remove thin disk stars and the same distance cut at $<10$~kpc \citep[with distances from][]{BailerJones_distances}. Likewise, only giants with $\log g<3$ and $T_\textrm{eff}<5200$ are included in the catalog; note that this catalog is itself built upon the catalog of giants provided from the BP/RP spectra by \citet{andrae_metallicities}. We use only the recommended high-reliability, low extinction sample of stars with a selection for $\textrm{E}(\textrm{B}-\textrm{V})<0.2$ \citep[with the $\textrm{E}(\textrm{B}-\textrm{V})$ values from][]{sfd_ebv}.

Our high-[N/O], high-[Al/Fe] sample from this dataset is constructed with the classification criteria recommended in \citet{Kane_2024}: $\mu_{\textrm{[N/O]}}-0.19\times\sigma_{\textrm{[N/O]}}>0.65$, $\mu_{\textrm{[Al/Fe]}}-\sigma_{\textrm{[Al/Fe]}}>0.$, $\mu_{T_\mathrm{eff}}<5000$, and $\mu_{\textrm{[Fe/H]}}>-2.0$ (with these latter two cuts used to select stars that are well-represented in the training data). In validation testing with APOGEE data, these selection criteria identify true 2P-type stars with a $\sim7\%$ false positive rate, although many of the false positives in the validation testing were close to the true boundary and may actually be 2P stars. These cuts yield 577 high-[N/O] candidates in the field, for which the median stellar parameters are $T_\mathrm{eff}=5859~\textrm{K}$, $\log g =1.72$, and $\textrm{[Fe/H]}=-1.41$. Our chemically typical background sample from this dataset is selected with the cut $\mu_{\textrm{[N/O]}}-0.19\times\sigma_{\textrm{[N/O]}}<0.4$; much as with the high-[N/O] selection, this somewhat more strict cut is intended to fully exclude high-[N/O] stars by accounting for errors in the neural network inferred abundances, which are naturally higher than the APOGEE errors. The same $\mu_{T_\mathrm{eff}}<5000$ and $\mu_{\textrm{[Fe/H]}}>-2.0$ cuts are enforced on the background sample as the high-[N/O] candidates for consistency. This background sample contains 57\,470 stars.

\subsubsection{The Galactic Globular Clusters}
\label{subsec:GC_data}

We use the set of surviving GCs provided in Holger Baumgardt's GC catalog\footnote{\url{https://people.smp.uq.edu.au/HolgerBaumgardt/globular/}}. The catalog includes GC motions and distances from \citet{vasiliev_gc} and masses from \citet{Baumgardt_2017,Baumgardt_Hilker_2018,Baumgardt_2023}. GC metallicities are adopted from \citet{Belokurov_2024_GCs}, where they are compiled from \citet{Harris_2010} and other literature sources. Here, we impose a metallicity cut of $\textrm{[Fe/H]}<-0.9$ on the GCs---slightly higher than the cut imposed on the stars to account for uncertainties in the stellar metallicities. We also remove the single GC with no recorded metallicity in the catalog, yielding 122 total GCs that we use. Notably, because of the faster chemical enrichment in the inner Galaxy, omitting GCs with $\textrm{[Fe/H]}>-0.9$ may remove some inner MW clusters that are not on strictly disk-like orbits. However, this cut is nonetheless consistent with those that we impose on the field stars and allows us to maintain a feasible sample size of field stars for our computations; future work could consider extending the method to more metal-rich GCs, particularly in the inner MW.

\subsection{\texorpdfstring{Connecting High-[N/O] Field Stars and GCs in $E-L_z$ Space}{Connecting High-[N/O] Field Stars and GCs in E-Lz Space}}
\label{subsec:2D_method}

Our goal is to identify stars with 2P-type chemistry in the Galactic field with orbital properties consistent with their having originated in surviving GCs; these are ``escapee" stars from their birth cluster. As such, we examine three groups of objects in $E-L_z$ space: 1.) globular clusters, 2.) high-[N/O], high-[Al/Fe] stars in the Galactic field \citep[from both APOGEE and the candidates in][as identified by the criteria discussed in Sections~\ref{subsec:APOGEE_data} and \ref{subsec:candidate-data}]{Kane_2024}, and 3.) the background sample of chemically typical field stars. In each case, we draw 10\,000 Monte Carlo samples per object from a normal distribution centered at the reported values of the radial velocities, proper motions in right ascension (RA) and declination, and distance with spreads equal to the reported uncertainties, with the correlation between the errors in proper motions in RA and Dec included. RA and Dec are the reported \textit{Gaia} DR3 values for all draws on the objects' errors. 

Each of the realizations of the observables for each star or GC are then used to calculate $E$ and $L_z$ using \texttt{AGAMA} \citep{AGAMA} in the Milky Way potential from \citet{McMillan2017}. We adopt \texttt{Astropy}'s \citep{astropy:2013,astropy:2018, astropy:2022} Galactocentric reference frame for our coordinates. These realizations of $E$ and $L_z$ are then assigned to one of a 500 by 500 bin grid in $E-L_z$ space, with the bins ranging from the $0.5^\textrm{th}$ or $1^\textrm{st}$ to $99^\textrm{th}$ percentile of all of the clusters' aggregate energy or angular momentum PDFs, respectively (the extended lower bound on energy is necessary not to exclude a very low energy cluster altogether).

By propagating the errors into calculations of the star or cluster's energy and angular momentum, we produce 2D probability density functions (PDFs) of each GC or star in the $E-L_{z}$ plane, $P(E,L_z)$. We show the PDFs of the GCs and high-[N/O] stars in $E-L_z$ space in Fig.~\ref{fig:ELz_hist}. Note that the PDFs of the clusters are much better constrained than those of the individual field stars, which is to be expected as the measurements for entire clusters are themselves much better constrained as each such measurement is based on a large number of cluster stars. Note also that the BP/RP stars' errors are generally better constrained than those for the APOGEE targets, which arises from the fact that the BP/RP stars have generally lower proper motion errors due to their presence at higher ecliptic latitudes than the APOGEE targets. However, the APOGEE stars' \texttt{AstroNN} \citet{astroNN-Leung_Bovy_2019} distances are slightly better constrained than the \citet{BailerJones_distances} geometric distances used for the BP/RP sample. The overall distributions of the APOGEE and BP/RP N-rich stars also differ, with the APOGEE stars extending to somewhat lower energies, likely because APOGEE makes its observations in infrared \citep{Wilson_APOGEE_spectrographs} and can thus more easily observe towards the Galactic center and plane.

\begin{figure*}
	\includegraphics[width=7in, alt={The axes are the same z-component of angular momentum and energy as are used in Figure 1. The high probability excess GCs are NGC 6362, NGC 6642, and NGC 6121. They all have orbital energies well below Solar and relatively small, well constrained PDFs. The overlap, or share space in the E-Lz plane, with 5 or more PDFs of APOGEE N-rich stars, which appear as less well-constrained streaks. By contrast, the low probability excess GCs, which are AM 4, NGC 6715, and IC 4499, all have energies well above Solar. NGC 6715 and IC 4499 have only slightly less well-constrained PDFs than the high probability GCs, but NGC 6715 only overlaps with the low probability tails of a few N-rich star PDFs, and IC 4499 overlaps with one single N-rich star PDF. AM 4 has a large, puffy, unconstrained PDF "cloud" and thus overlaps with many N-rich star PDFs, but presumably the large size of its PDF means that it also overlaps with an even more substantial number of comparison stars, which are not shown here.}]{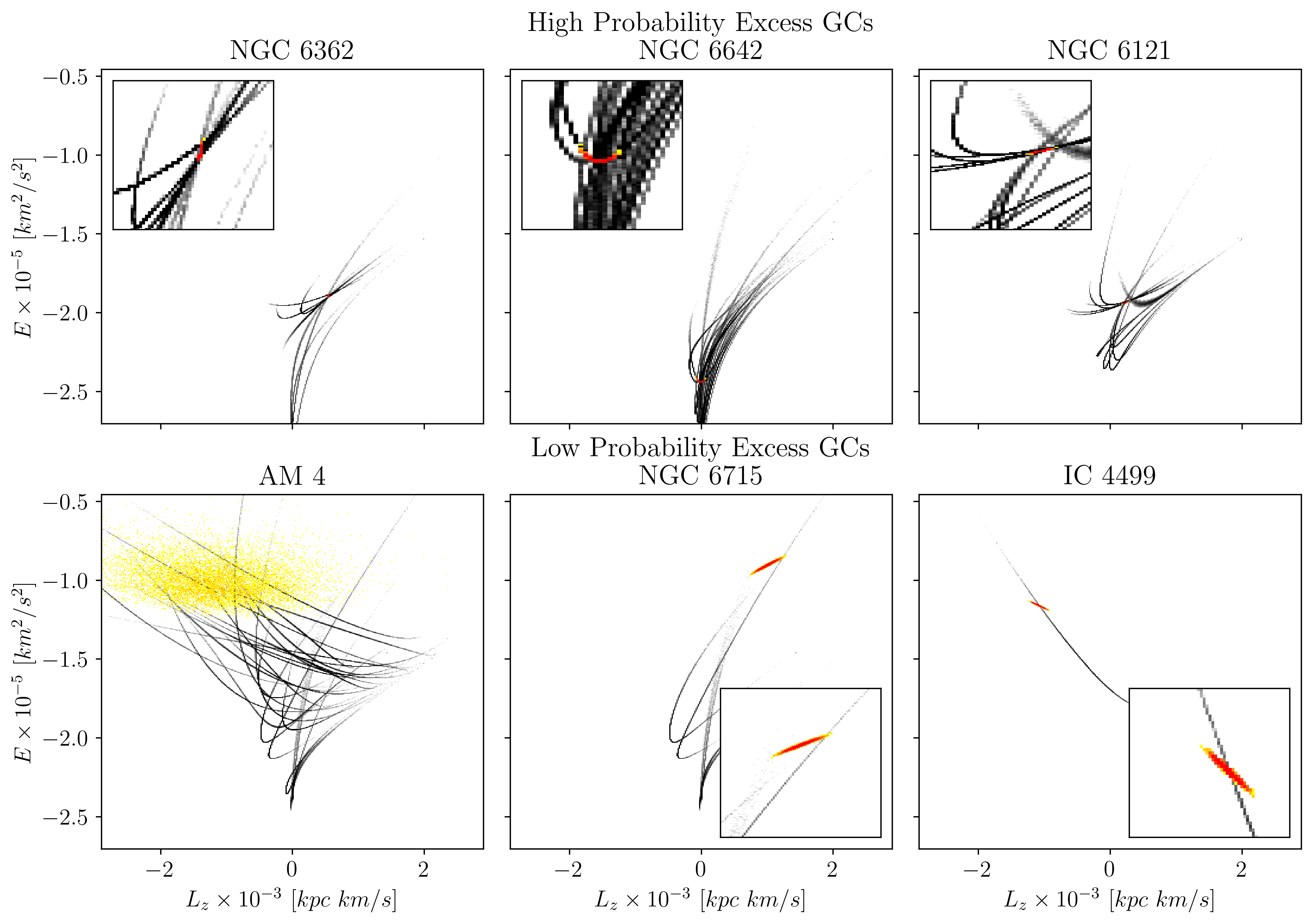}
    \caption{In each panel, the PDF of an individual cluster in $E-L_z$ space is shown in orange. Included in the same panel are the PDFs of every high-[N/O] field star from APOGEE with non-zero association $\epsilon_{c,~s}$ with that cluster. Top row: Clusters with high probability of association with N-rich field stars in $E-L_z$ space. Bottom: Clusters with low probability and low probability excess of association with N-rich field stars. Several panels also include zoom-in insets of the region of the figure containing the relevant GC for easier visualization.}
    \label{fig:example_overlap}
\end{figure*}

To quantify the associations between stars and GCs, we then take the product of the GC PDFs with the PDFs of the high-[N/O] candidates and integrate those products to get the probability per star that it is associated with a GC in the $E-L_{z}$ plane (or, phrased differently, to quantify the degree to which the PDFs of the star and GC overlap in $E-L_{z}$ space). The same calculation of taking the integrated products of the star and GC PDFs is performed separately for the comparison sample of stars. The association, which we label $\epsilon_{c,s}$ between the $c$th GC and the $s$th high-[N/O] field star is thus:
\begin{equation}
    \epsilon_{c,~s} =  \int P_c(E,L_z) P_s(E,L_z) \,dE \, dL_z
\end{equation}
where $P_c$ and $P_s$ are the probability distribution of the cluster and the star, respectively, in $E-L_z$ space. $\epsilon_{c,s}$ is thus nonzero if GC $c$ and star $s$'s PDFs have some degree of overlap. We can also look at the association of a single GC in aggregate with all of the high-[N/O] field stars, $\epsilon_{c}$:
\begin{equation}
    \epsilon_{c} = \sum_{s} \int P_c(E,L_z) P_s(E,L_z) \,dE \, dL_z / s_\mathrm{tot}
\end{equation}
where $s_{tot}$ is the total number of stars $s$. This concept is illustrated in Fig.~\ref{fig:example_overlap}, wherein high probability ($\epsilon_{c}$) clusters (top row) overlap with many stars' PDFs, and low probability clusters overlap with few.

In the case of the aggregate probability for a single cluster $\epsilon_{c}$, we can scale the association of the GC and the high-[N/O] field stars by the association between the same GC and the background (chemically typical) halo stars to determine a "probability excess," or the degree to which the GC shares similar $E$ and $L_z$ with the N-rich field stars as compared to the halo as a whole. This probability excess for GC $c$ is thus simply defined as $\epsilon_{c,~\textrm{excess}}$:
\begin{equation}
\label{eq:prob_excess_ELz}
    \epsilon_{c,~\textrm{excess}} = \frac{\sum_\textrm{N-rich} \int P_c(E,L_z) P_\textrm{N-rich}(E,L_z) \,dE \, dL_z / N_\textrm{N-rich}}{\int P_c(E,L_z) \sum_\textrm{comp} P_\textrm{comp}(E,L_z) \,dE \, dL_z / N_\textrm{comp}} 
\end{equation}
Thus, in cases where a GC has equal associations with the comparison set of chemically typical stars as with the high-[N/O] halo stars, $\epsilon_{c,~\textrm{excess}}=1$. Similarly, $\epsilon_{c,~\textrm{excess}}>1$ if the association between the GC $c$ and the N-rich stars are greater than that with the comparison sample. This scaling to the comparison sample is only possible for comparing the association between a GC and two populations as a whole and is not as feasible for any individual star, as constructing a reasonable "comparison" for a single star is not straightforward. Using the probability excess rather than $\epsilon_{c}$ alone is useful not only because it allows us to trace differences between the distributions of the high-[N/O] stars and the chemically typical halo relative to the GCs but also because it helps account for the effect of uncertainties, especially for the clusters' $E-L_z$ distributions. For instance, AM-4 has very uncertain energy and angular momentum constraints due to large errors in its proper motions and radial velocity measurements, resulting in a large PDF that intersects with many high-[N/O] star PDFs (see the lower left panel of Fig.~\ref{fig:example_overlap}). Nonetheless, AM-4 is classified as a cluster with low probability excess because that large spread in $E$ and $L_z$ likewise produces large overlap with many stars in the comparison sample. We have checked the final probability excesses and confirmed that they do not appear to correlate with the errors on the observables (and thus the spread on the PDFs).

Note that in the case of the comparison sample, the association is calculated between a single GC and the comparison sample as a whole rather than for each star individually (i.e., the integrated product is taken between a single GC PDF and the sum of all of the comparison star PDFs, see the denominator of Equation~\eqref{eq:prob_excess_ELz}). Mathematically, this is equivalent to the individual calculation of association per star per GC that is performed for the high-[N/O] field stars but is much less computationally expensive. Given that we are interested only in the bulk properties of the comparison sample rather than the individual stars, this is a reasonable step to take. The computation is performed separately for the APOGEE and BP/RP stars, which is necessary given that they have different comparison sets.

As a check on the method, we split the chemically typical stars into a comparison and validation sample such that the validation sample replaces the N-rich stars in the numerator of Eq.~\ref{eq:prob_excess_ELz}. Thus, the validation probability excess should be approximately 1, since that excess is scaled to the comparison set, which is itself a subset of the overall sample of field stars from which the validation set is also chosen. To divide the background (chemically typical field stars) sample into a comparison and validation set, we randomly divide the background set into 80\% and 20\%, respectively. This process is repeated 10 times to produce 10 comparison and 10 validation sets, for which the probability per GC $\epsilon_c$ is calculated individually. The final comparison and validation $\epsilon_c$ we use is then taken as the average of the result from the ten samples. We also tested the validation set without resampling and nonetheless recover a similar result of $\epsilon_\textrm{excess, validation}\approx1$ across the clusters, but there is much more noise, particularly in the APOGEE dataset where there are substantially fewer stars. The step of resampling the validation and comparison dataset thus helps to make the results more easily interpretable by reducing noise in the background dataset.

An important caveat to note is that it is not reasonable to expect \emph{all} high-[N/O] field stars to be associated with an existing GC via their energy and angular momentum. Some cluster "escapees" will have had their orbits sufficiently changed from their cluster of origin that we can no longer connect them via their orbital properties; others will have originated in clusters that have since dissolved entirely and thus will not exist in the GC catalog \citep[both of these possibilities are discussed in][]{Savino_Posti_2019}. Further complicating matters, some of these entirely dissolved GCs may have had similar orbital characteristics to present-day surviving clusters. These nuances further motivates looking at the association of high-[N/O] stars with existing GCs only as an excess as compared to a selection of other "background" stars. Even with these caveats in mind, if the high-[N/O] stars selected \emph{do}, in fact, exhibit such peculiar chemistry due to an origin in GCs, then we should expect them as a population to be more likely to be associated with GCs than an approximately random sample of field stars with no such selection for a GC origin.

\subsection{Extending the Method to Other Spaces}
\label{subsec:3D4D_method}

Although the $E-L_z$ plane has been routinely used to identify clustering and substructure, which motivates its inclusion in this work, it suffers from several challenges. First, by its nature the plane is crowded at low energies, with many stars and GC PDFs overlapping substantially (as can be observed in Fig.~\ref{fig:ELz_hist}). Inconveniently, this is the region of $E-L_z$ space where the most high-[N/O] field stars are located. Second, there is of course no chemical information included in this space, meaning the high-[N/O] stars need not share, for instance, a metallicity with the GCs to which they are ``tagged,” which would be a sensible expectation for stars that were truly formed in that GC. We attempt to address both of these issues by expanding the method to a modified three-dimensional space of energy, rescaled $L_z$, and metallicity and also to a four-dimensional space of the three orbital actions ($\vec{J}$) and metallicity.

\begin{figure*}
	\includegraphics[width=7in, alt={In this space of Lz re-scaled by the angular momentum of a circular orbit at the given energy on the horizontal axis and orbital energy on the vertical axis, the values of re-scaled angular momentum are "stretched" at lower energies and thus are not as restricted as they are in the typical E-Lz plane, as in Figure 1. This stratching effect means that PDFs do not overlap as much at low energies, especially the GC PDFs; there are sufficient numbers and concentrations of the N-rich field stars at low energies that their PDFs nonetheles do intersect. The GC PDFs are again the most well-constrained, although the stretching effect means that the lowest energy GCs have PDFs which appear more like "strings" than "clouds." The APOGEE stars' PDFs are again less constrained than the BP/RP stars' or the GCs' and thus appear as long "strings" across the plane. Again, the N-rich field stars are mostly clustered at lower energies as compared to the GCs, but the BP/RP field stars do not extend to as low energies as the APOGEE stars.}]{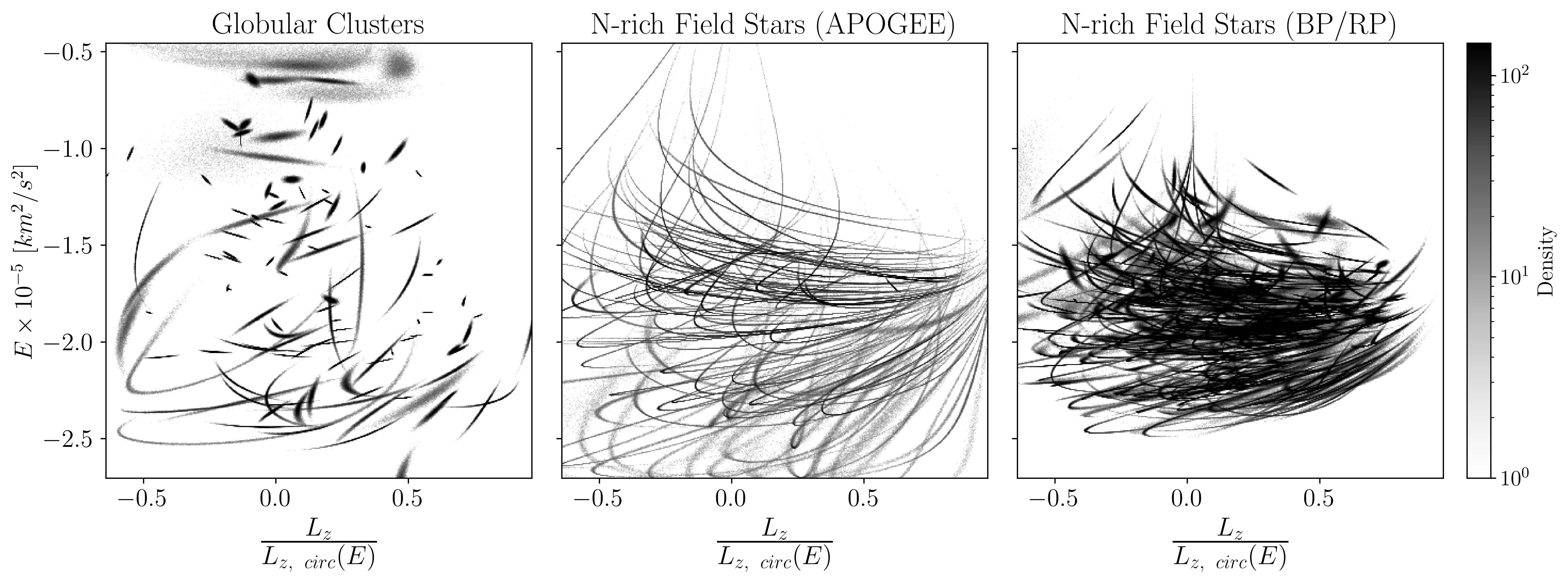}
    \caption{The 2D histograms of the distributions of the draws from the errors of the GCs (left), the high-[N/O] field stars from APOGEE (middle), and the high-[N/O] stars from the BP/RP sample (right) in $E-L_z/L_{z,~\textrm{circ}}(E)$ plane. Note that the third dimension used in this clustering space, [Fe/H], is not visualized here.}
    \label{fig:ELz_scaled_hist}
\end{figure*}

To extend our method to a three-dimensional space of energy, rescaled $L_z$, and metallicity, the actual procedure is almost identical to that outlined in Section~\ref{subsec:2D_method}, with 10\,000 draws being randomly selected from the observables of the GCs and field stars in order to produce a PDF for each object. These samples from the astrometry are again used to calculate $E$ and $L_z$ in the \citet{McMillan2017} potential using \texttt{AGAMA}, but this time we also calculate $L_{z,~\textrm{circ}}(E)$, the $z$-component of the angular momentum of a circular orbit at the given energy $E$. Instead of using $L_z$, we then calculate a rescaled angular momentum $L_z/L_{z,~\textrm{circ}}(E)$. This has the effect of ``flattening out" the energy-angular momentum plane, as the values of $L_z/L_{z,~\textrm{circ}}(E)$ are not as restricted as $L_z$ at low values of $E$. The PDFs of the GCs and the high-[N/O] field stars from both APOGEE and the BP/RP sample in the $E-L_z/L_{z,~\textrm{circ}}(E)$ plane are shown in Fig.~\ref{fig:ELz_scaled_hist}. As can be observed in the figure, there is far less ``crowding" of the PDFs at lower energies than in Fig.~\ref{fig:ELz_hist}.

\begin{figure*}
	\includegraphics[width=7in, alt={The GCs' PDFs are depicted and appear as clouds; generally, the PDFs at higher action values are larger and those at lower action values are smaller and better constrained. The PDFs at lower values (close to 0 in all three projections) overlap noticeably, making it difficult to distinguish individual GCs in these regions. There is a noticeable concentration of GCs at around action values of 0 in all three coordinates, but there is also an extent of GC PDFs into positive values of J-r and J-z and both positive and negative values of J-phi.}]{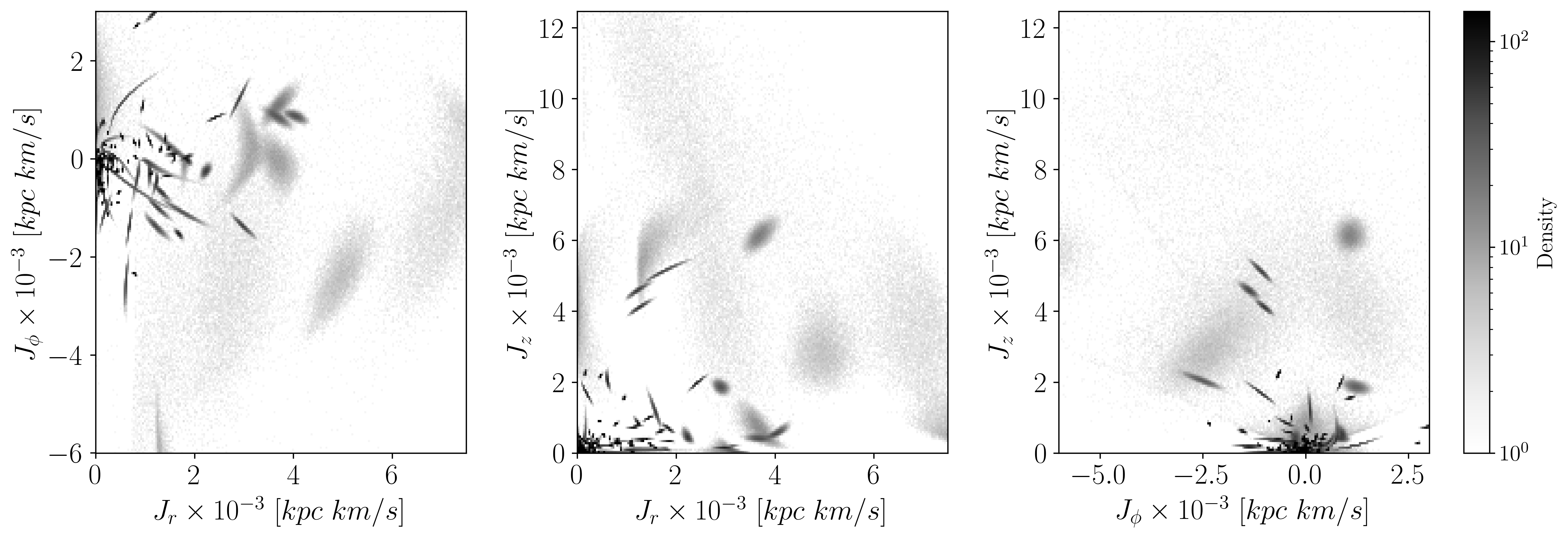}
    \caption{The 2D histograms of the distributions of the draws from the errors of the GCs in the $J_\phi-J_r$, $J_z-J_r$, and $J_z-J_\phi$ planes. Note that the fourth dimension used in this clustering space, [Fe/H], is not visualized here.}
    \label{fig:actions_hist}
\end{figure*}

Regarding metallicity, the third dimension used in this space, GC [Fe/H] abundances are taken from \citet{Belokurov_2024_GCs} (see Section~\ref{subsec:GC_data}), with most clusters being assigned a standard deviation in [Fe/H] of 0.05 dex \citep[approximately consistent with the median metallicity dispersion of 0.045~dex reported in][]{GC_metallicity_spreads}. However, some Galactic GCs, most famously $\omega$ Centauri, are known to host much larger [Fe/H] spreads than a typical cluster \citep[see the discussion of Type II GCs in][]{milone_multiple_pops_GCs}. Correspondingly, six GCs with more substantial Fe spreads ($\sigma_\mathrm{[Fe/H]}>0.1$) are set to the dispersions calculated in \citet{GC_metallicity_spreads} from spectroscopic measurements in the literature.
In particular, $\sigma_\textrm{[Fe/H]}$ for NGC 5139 ($\omega$ Cen) is set to 0.271 \citep[wherein the dispersion is calculated from][]{Willman_2012}; for NGC 5286, $\sigma_\textrm{[Fe/H]}=0.103$ \citep[from][]{Marino_2015}; for NGC 6205 (M13), $\sigma_\textrm{[Fe/H]}=0.101$ \citep{Masseron_2019}; for NGC 6273 (M19), $\sigma_\textrm{[Fe/H]}=0.161$ \citep{Johnson_2017,Johnson_2015}; for NGC 6656 (M22), $\sigma_\textrm{[Fe/H]}=0.132$ \citep{Marino_2011}; and for NGC 6715 (M54), $\sigma_\textrm{[Fe/H]}=0.183$ \citep{Carretta_2010}.

For more typical (Type I) GCs, a spread of 0.05 dex in either direction from the reported value of the cluster’s [Fe/H] may in fact be too high for some clusters, especially given that the high-[N/O] stars are second generation stars, which are routinely observed to have small iron spreads \citep[metallicity spreads in typical GCs, especially among their first generation stars, is still something of an emerging topic; for multiple views on the topic, see][]{Marino_2019,Marino_2023,Legnardi_2022,Carretta_1G_metallicity,Carretta_1G_metallicity_2}, but this is done intentionally to account for the fact that there may be systematic offsets and dispersions between the GC catalog metallicities and those for the stars (especially those derived with the very low resolution BP/RP spectra). Nevertheless, the code will be made public and can easily be re-run with different cluster metallicity spreads, as desired.

For APOGEE stars, we use the reported value of [Fe/H] as the center of the distribution from which we make our draws and the reported error on [Fe/H] as the spread; for the BP/RP sample, the network's inferred [Fe/H] ($\mu_\textrm{[Fe/H]}$) for each star is taken as the center of each distribution and the inferred standard deviation on that prediction ($\sigma_\textrm{[Fe/H]}$) is taken as the spread. Naturally, given both the resolution of the spectra and the method of abundance derivation, the uncertainties on the BP/RP metallicities are much higher than those from APOGEE; the median [Fe/H] error of the APOGEE high-[N/O] stars is $0.013$, whereas for the BP/RP high-[N/O] candidates it is $0.26$. From these metallicities and their associated spreads, 10\,000 normally distributed samples in [Fe/H] are generated per object, analogous to the kinematic sampling.

The result of these calculations is now a 3D PDF for each object, or in aggregate for the comparison and validation samples, written as $P(E,L_z/L_{z,~\textrm{circ}}(E),\textrm{[Fe/H]})$. The samples are separated into 400 bins each in the $E$ and $L_z/L_{z,~\textrm{circ}}(E)$ dimensions, with the bins ranging from the $0.5$th to the $99$th or the $0.5$th to the $99.5$th percentile of the GCs' aggregate samples in each dimension, respectively. The [Fe/H] dimension is grouped into only 30 bins, corresponding to a resolution of approximately 0.05~dex. Note that the number of bins in the energy and rescaled angular momentum dimensions is reduced from 500 in Section~\ref{subsec:2D_method} to account for the additional bins added in [Fe/H]; in reality, however, this choice of binning has very little effect on the ultimate probability excesses. The result is a 3D histogram for each object, constructed from $400\times400\times30$ bins.

If we define the PDF of an object in this 3D case as:
\begin{equation}
    P(E, \frac{L_z}{L_{z,~\textrm{circ}}(E)},\mathrm{[Fe/H]})\equiv  P(3\textrm{D})
\end{equation}
then the probability excess, 
$\epsilon_{c,~\textrm{excess}}$, becomes:
\begin{align}
    &\epsilon_{c,~\textrm{excess}} = \label{eq:prob_excess_3D}  \\
    &\frac{\sum_\textrm{N-rich} \int P_c(3\textrm{D}) P_\textrm{N-rich}(3\textrm{D}) \,dE \,d\frac{L_z}{L_{z,~\textrm{circ}}(E)} \,d\mathrm{[Fe/H]} / N_\textrm{N-rich}}{\int P_c(3\textrm{D}) \sum_\textrm{comp} P_\textrm{comp}(3\textrm{D}) \,dE \,d\frac{L_z}{L_{z,~\textrm{circ}}(E)} \,d\mathrm{[Fe/H]} / N_\textrm{comp}} \nonumber 
\end{align}
As in the 2D case, $\epsilon_{c,~\textrm{excess, validation}}$ has an identical form with the validation sample in aggregate replacing the high-[N/O] stars in the numerator. 

To explore associations in the four-dimensional action-metallicity space, the astrometry is now used to calculate the actions, $\Vec{J}=(J_r,J_\phi,J_z)$, using \texttt{AGAMA} (again in the \citet{McMillan2017} potential) rather than energy and $L_z$. \texttt{AGAMA} calculates $J_\phi$ as the $L_z$ component of angular momentum, while the other action components $J_r$ and $J_z$ are approximated via the St\"ackel fudge \citep{Binney_2012}. Metallicities and their standard deviations for both the stars and GCs are adopted via the same method as was used in the 3D case.

The 4D PDF is then constructed from the 10\,000 samples of the actions and metallicities for each object, $P(\Vec{J},\mathrm{[Fe/H]})$ (or a single PDF for all stars for the comparison and validation sets, as in Section~\ref{subsec:2D_method}), with three dimensions being the components of $\Vec{J}$ and the fourth being [Fe/H]. Each action dimension has 200 bins, and the metallicities are again grouped into 30 bins. The bins span from $0$ to the $99^\textrm{th}$ percentile of the GCs' sampled values in $J_r$ and $J_z$ and from the $1^\textrm{st}$ to $99^\textrm{th}$ percentile values in $J_\phi$ and [Fe/H]. 
We note that these bins in the actions are not sufficient to resolve the PDFs of some of the more well-constrained GCs. However, given the challenges of actually tagging a field star to a \emph{single} cluster, which is discussed in more detail in Sections~\ref{sec:results}~and~\ref{sec:discussion}, and that the goal of this work is to look at the bulk properties of the high-[N/O] stars in relation to the Galactic globular clusters, we accept this decreased precision in resolving the PDFs. Naturally, the 4D PDF of actions and metallicities is not readily visualizable, so we provide 2D projections of the GCs' action space in Fig~\ref{fig:actions_hist} to illustrate the general distribution of clusters in the space as well as the effect of our binning.

The probability excess, $\epsilon_{c,~\textrm{excess}}$, in the 4D case becomes:
\begin{align}
    &\epsilon_{c,~\textrm{excess}} = \label{eq:prob_excess_4D} \\
    &\frac{\sum_\textrm{N-rich} \int P_c(\Vec{J},[Fe/H]) P_\textrm{N-rich}(\Vec{J},[Fe/H]) \,d\Vec{J} \, d[Fe/H] / N_\textrm{N-rich}}{\int P_c(\Vec{J},[Fe/H]) \sum_\textrm{comp} P_\textrm{comp}(\Vec{J},[Fe/H]) \,d\Vec{J} \, d[Fe/H] / N_\textrm{comp}} \nonumber 
\end{align}

As we discuss in Section~\ref{sec:intro}, several previous works have tagged field N-rich stars to GCs based upon their actions and metallicities \citep{Savino_Posti_2019,Hanke2020_method,Xu2024_method}. Aside from the different datasets used here and the fact that we look at associations using other integrals of motion in addition to actions, our work is distinct from these past studies in two significant ways. First, rather than estimating the uncertainties on each action individually from the Monte Carlo samples \citep{Savino_Posti_2019} or using a Kernel Density Estimator (KDE) to construct the PDF of the difference between the stars' and GCs' properties \citep{Hanke2020_method, Xu2024_method}, we bin the data in an n-dimensional histogram (here, where n is either 2, 3 or 4) to represent the PDFs of each object and compare those representations. Second and perhaps more importantly, the use of a comparison dataset allows us to set a threshold for association of N-rich field stars with GCs, such that when $\epsilon_{c,~\textrm{excess}}>1$, a cluster can be considered a likely candidate for association with high-[N/O] field stars. This enables us to study the bulk properties of the clusters that likely deposited the N-rich stars into the halo, and means that this work is less focused on individually tagging any individual star to the cluster in which it originated (see Sections~\ref{sec:results} and \ref{sec:discussion} for further discussion).

\section{Results}
\label{sec:results}

\begin{figure*}
	\includegraphics[width=2\columnwidth, alt={At energies slightly lower than Solar, the median probability excess increases to approximately 1.75 to 2; at lower energies, the median probability excess is <1. For clusters with an initial mass greater than about 1,000,000 Solar masses, the median probability excess increases above 1 to between 2.5 and 4. The median probability excess also transitions from <1 to >1 for GCs which have lost more than about 1,000,000 Solar masses. Median probability excesses calculated in the energy-angular momentum plane increase above 1 at slightly higher energies and and lower GC initial masses and mass losses than those calculated with the other two methods.}]{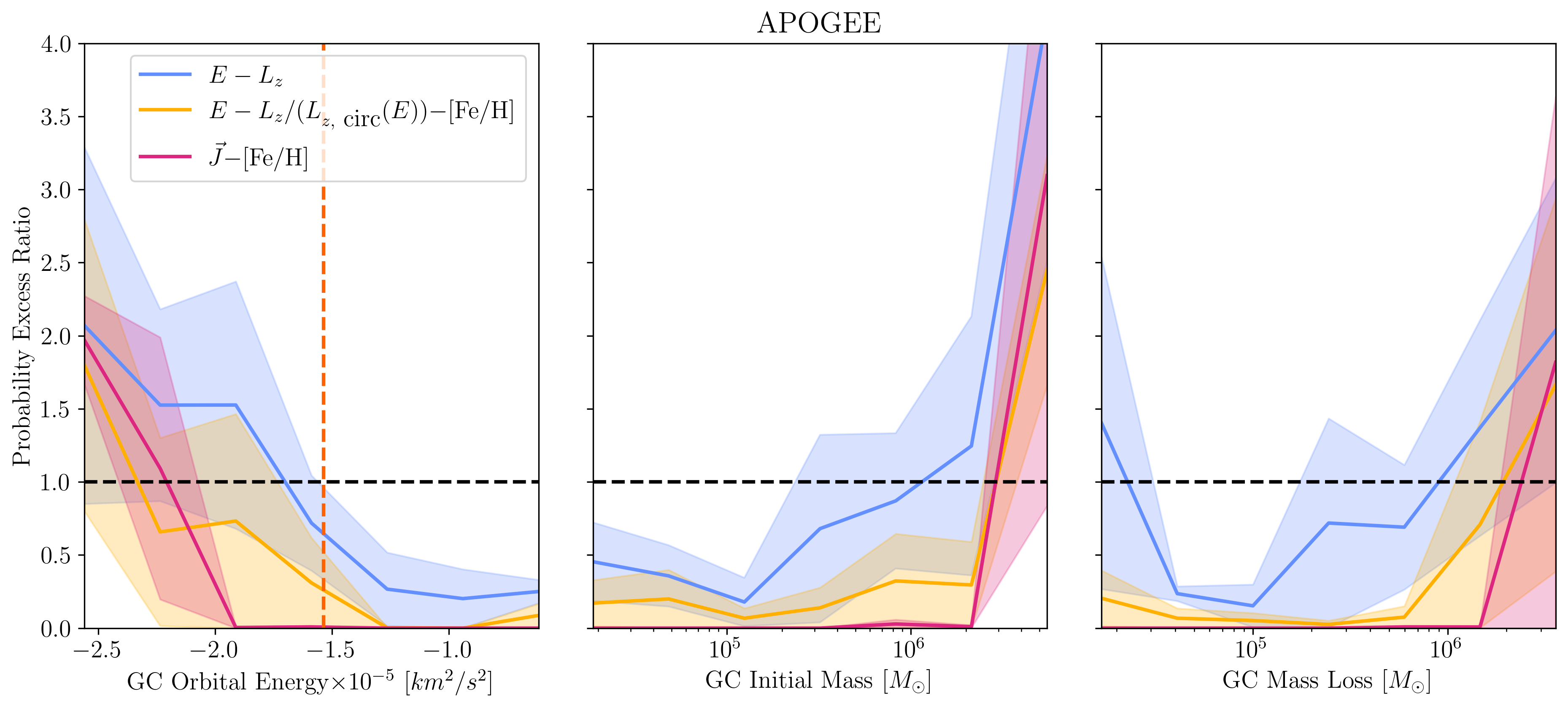}
    \caption{For the APOGEE data, the probability excess of high-[N/O] field stars being associated with GCs relative to the comparison sample of chemically typical field stars (see Eqs.~\ref{eq:prob_excess_ELz}, \ref{eq:prob_excess_3D} and \ref{eq:prob_excess_4D}). The lines denote the median probability excess as a function of GC orbital energy in the \citet{McMillan2017} potential (left), initial GC mass (middle), and mass lost by the GC over its lifetime ($\Delta M$, right). The shading shows the median absolute deviation in each bin. The results are shown separately for associations in the $E-L_z$ plane (blue), the 3D $E-L_z/L_{z,~\textrm{circ}}(E)-\textrm{[Fe/H]}$ space (yellow), and the 4D action-metallicity space (pink). The black dashed line marks a probability excess of 1, representative of an equal association between the high-[N/O] and comparison field stars with the GCs. In the left panel, the vertical orange dashed line marks the Solar energy, approximately \mbox{$-1.54\times10^5~\mathrm{km}^2\,\mathrm{s}^{-2}$}.}
    \label{fig:APOGEE_trends}
\end{figure*}

\begin{figure*}
	\includegraphics[width=2\columnwidth, alt={The trends in median probability excess here, for the BP/RP stars, are overall very similar to those calculcated with the APOGEE stars, described in Figure 5. At energies slightly lower than Solar, the median probability excess increases from <1 to 2, although at the lowest energies the median excess drops to the equal association line at 1. For clusters with an initial mass greater than about 500,000 Solar masses, the median probability excess increases above 1 to between 1.5 and 3. The median probability excess also transitions from <1 to >1 for GCs which have lost more than about 500,000 Solar masses. As compared to the probability excesses calculcated with the APOGEE stars, there is even more consistency between the median trends from the three different clustering spaces for the BP/RP stars, particularly regarding at which values GC energy, initial mass, and mass loss the probability excess crosses the equal association line at 1.}]{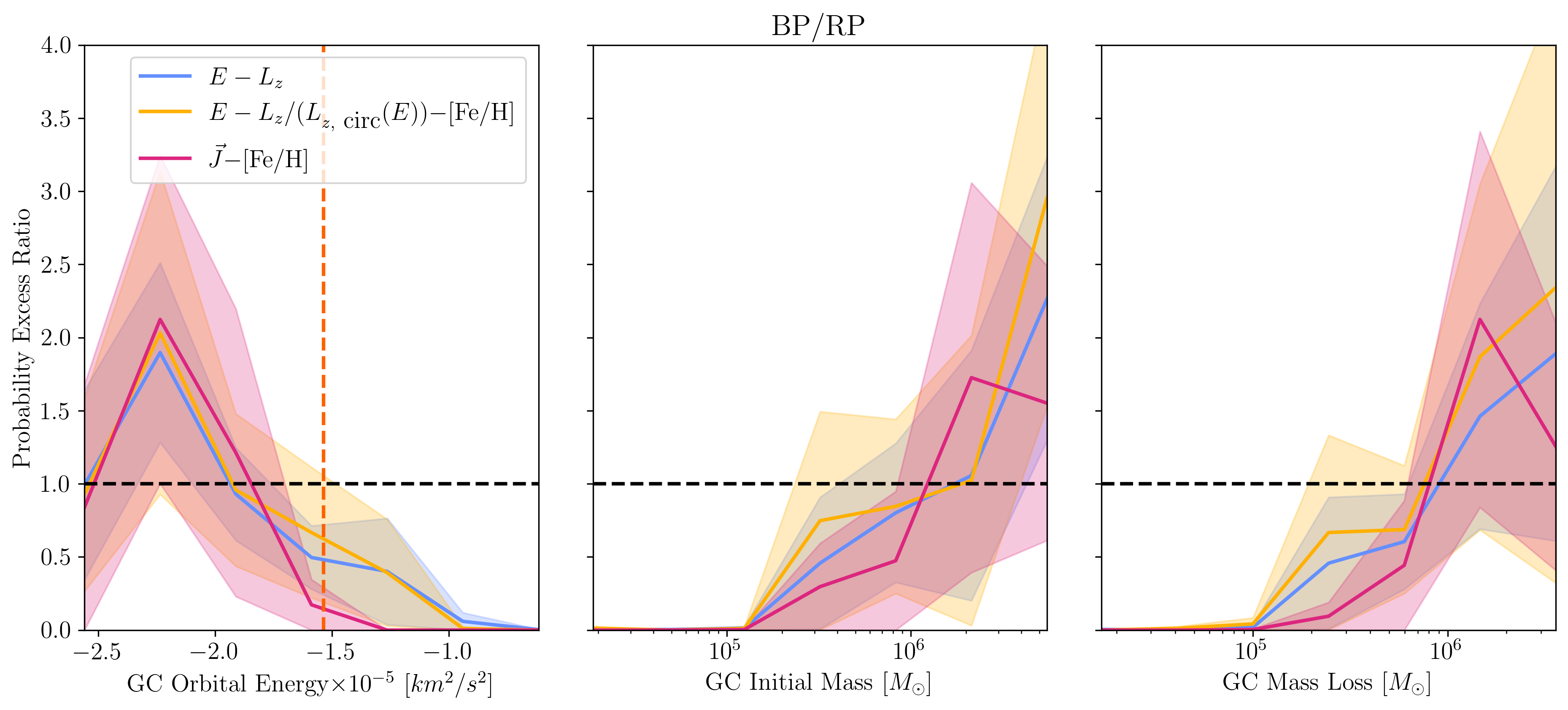}
    \caption{As in Fig.~\ref{fig:APOGEE_trends}, the probability excess of association of the high-[N/O] field stars as a function of GC orbital energy (left), initial GC mass (middle), and GC mass loss (right), now for the BP/RP sample. Again as before, the lines indicate the median excess per GC in each bin, and the shading shows the median absolute deviation. The results are again separated for the $E-L_z$ plane (blue), the 3D $E-L_z/L_{z,~\textrm{circ}}(E)-\textrm{[Fe/H]}$ space (yellow), and the 4D action-metallicity space (pink). The black dashed line marks a probability excess of 1, and in the left panel, the vertical orange dashed line marks the Solar energy.}
    \label{fig:BPRP_trends}
\end{figure*}

\begin{figure*}
	\includegraphics[width=5.5in, alt={The figure shows histograms of probability excesses per cluster split by accreted and in situ GCs, separately for the APOGEE and BP/RP results. Almost all of the GCs classified as accreted have probability excesses below 1 in both APOGEE and the BP/RP data, although about 4 accreted GCs have probability excesses greater than 1 from the BP/RP stars. By contrast, although many of the in situ-classified GCs have probability excesses less than 1, there nonetheless exists a substantial population with excesses greater as high as 4 to 6, which is the axis limit of this figure. There are even more such high probability excess in situ GCs from the BP/RP data than from the APOGEE data.}]{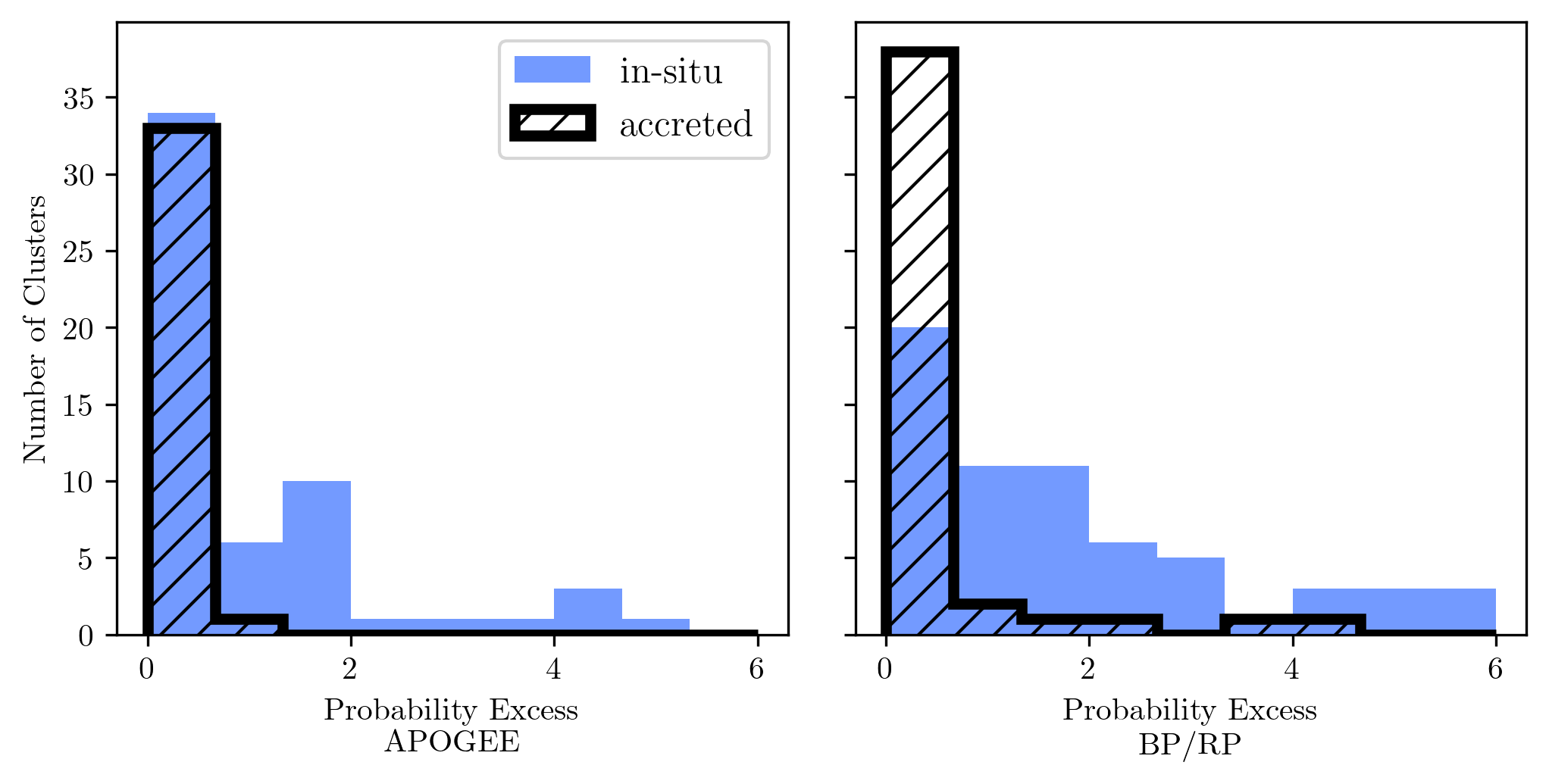}
    \caption{The probability excess per cluster for association with high-[N/O] field stars from APOGEE (left) and the BP/RP sample (right) as calculated from the actions, $\Vec{J}$, and metallicities, [Fe/H] (Section~\ref{subsec:3D4D_method}. The clusters are split into accreted and \textit{in situ} groups per the classifications from \citet{Belokurov_2024_GCs}, represented by the black hatched and blue histograms, respectively.}
    \label{fig:insitu_accreted_4D}
\end{figure*}

\begin{figure}
	\includegraphics[width=\columnwidth, alt={At low energies, many stars and GCs have nonzero scaled probabilities on the order of 1 or greater. For any single star, there are typically multiple clusters for which it has nonzero probabilities, and likewise, the lowest energy clsuters essentially all have nonzero scaled probabilities with many stars. The scaled probabilities are such that the lowest energy stars are tagged to the lowest energy clusters, and as the energies of the N-rich field stars increase, so do the energies of the GCs with which they share nonzero probabilities. At higher energies, the number of stars per cluster and clusters per star with nonzero decreases as compared to the lowest energies. There are only a few N-rich stars with energies greater than Solar, although about a third of the GCs have energies greater than Solar.}]{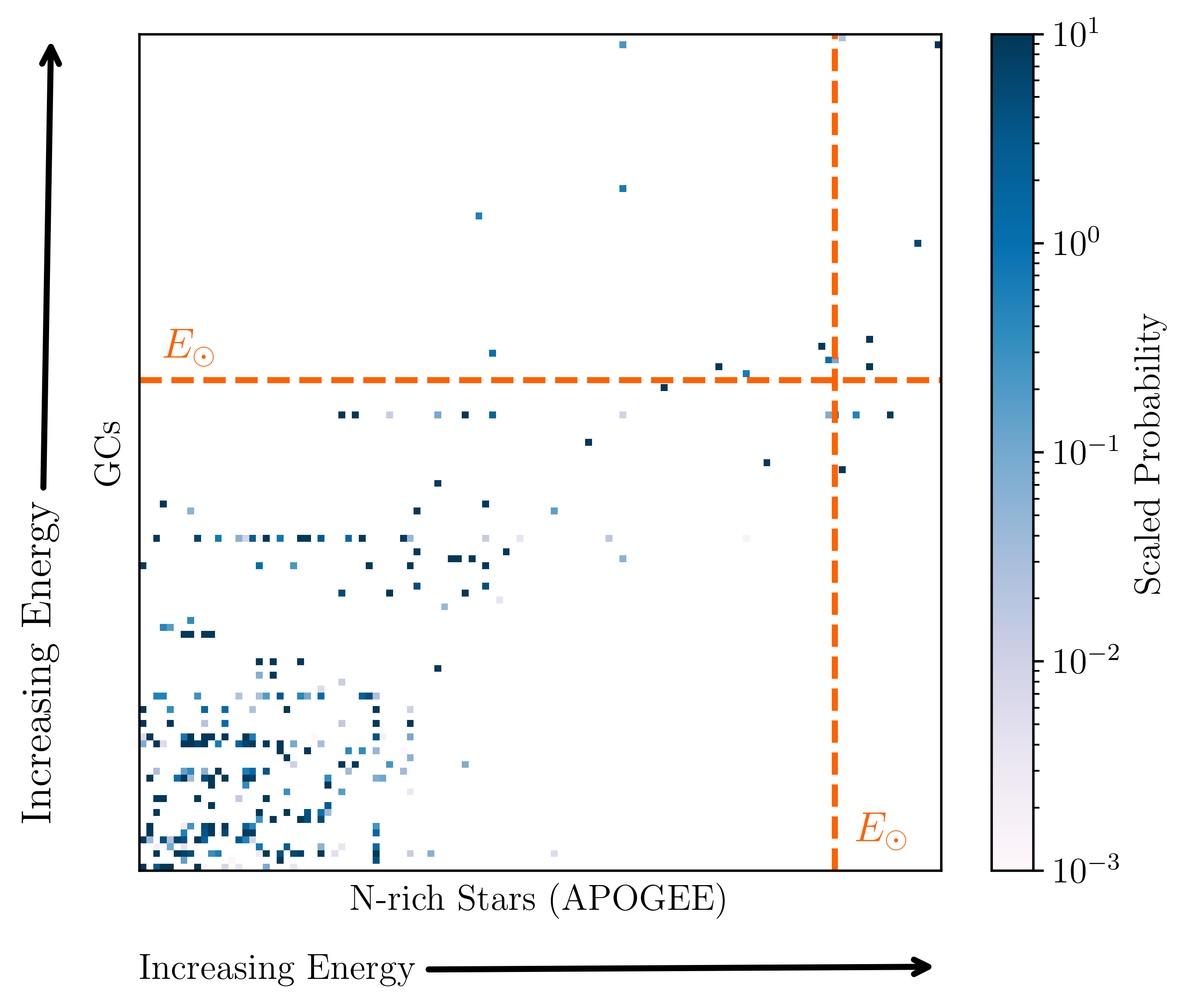}
    \caption{This figure is designed to be analogous to Fig.~17 of \citet{Hanke2020_method}. 
    It shows the product, per APOGEE N-rich star $s$ and per GC $c$, of the PDFs of the actions and metallicities for that star and cluster (Section~\ref{subsec:3D4D_method}):  
    $\protect\epsilon_{c,~s}=\int P_c(\vec{J},[Fe/H]) P_{\text{N-rich}}(\vec{J},[Fe/H]) \,d\vec{J} \, d[Fe/H]$.  
    Each product is scaled by the product of that GC's PDF with that of the comparison sample, $\protect\epsilon_{c,~\text{comp}}=\int P_c(\vec{J},[Fe/H]) \sum_{\text{comp}} P_{\text{comp}}(\vec{J},[Fe/H]) \,d\vec{J} \, d[Fe/H] / N_{\text{comp}}$.  
    Thus, each pixel represents a scaled probability $\protect\epsilon_{c,s}/\protect\epsilon_{c,\text{comp}}$.  
    The stars and GCs are all sorted by their energies, such that the lowest orbital energy N-rich stars are located on the left and the lowest energy clusters are located at the bottom.  
    As a reference, the orange dashed lines mark the approximate Solar energy in this potential \citep{McMillan2017}.}
    \label{fig:4D_prod_APOGEE}
\end{figure}

\begin{figure*}
	\includegraphics[width=2\columnwidth, alt={Many of the patterns in this figure are analagous to those in Figure 8, although, because the BP/RP sample is so much larger than the APOGEE N-rich sample,. In particular, at the lowest}]{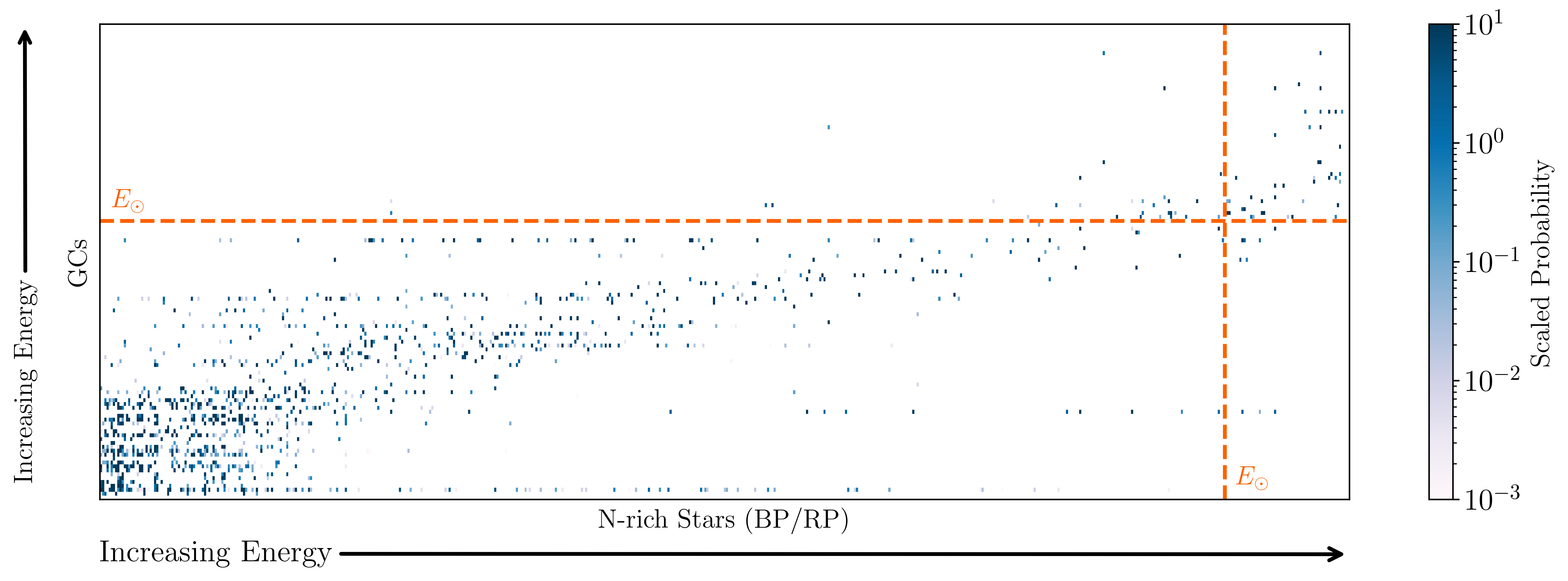}
    \caption{The same as Fig.~\ref{fig:4D_prod_APOGEE}, now for the BP/RP sample of high-[N/O] stars. Each pixel again represents the integrated products of each N-rich star $s$ and GC $c$ PDF in the 4D action-metallicity space, scaled by the comparison product for that cluster ($\protect\epsilon_{c,~s}/\protect\epsilon_{c,~\text{comp}}$). The stars and GCs are sorted by increasing orbital energy along the horizontal and vertical axis, respectively, with the orange dashed lines marking the Solar energy.}
    \label{fig:4D_prod_BPRP}
\end{figure*}

Utilizing the methods outlined in the previous section, we calculate for each GC a ``probability excess'' ($\epsilon_{c,~\mathrm{excess}}$) for association with the N-rich field stars relative to the association with the rest of the chemically typical halo. This excess is calculated separately in each of the three dynamical or chemo-dynamical spaces outlined, and the calculation is also performed separately for the APOGEE and BP/RP stars. In accordance with our goal of looking at the bulk chemodynamical associations of the high-[N/O] field stars with the Galactic GCs, we begin by examining the relationship of the probability excess per GC as a function of cluster properties. In particular, we examine the median probability excess relative to GC orbital energy, initial mass, and mass loss separately for the APOGEE stars (Fig.~\ref{fig:APOGEE_trends}) and the BP/RP-selected stars (Fig.~\ref{fig:BPRP_trends}). In both figures, the $\epsilon_c=1$ line is included to represent the approximate validation probability excess, or the value of $\epsilon_{c,~\textrm{excess}}$ denoting equal association between the GCs and the N-rich halo stars and the comparison sample. The true mean validation probability excesses for the $E-L_z$, $E-L_z/(L_{z,~\textrm{circ}}(E))-\textrm{[Fe/H]}$, $\vec{J}-\textrm{[Fe/H]}$ cases are $1.02$, $0.98$, and $1.03$, respectively, for APOGEE and $1.01$, $0.99$, and $1.02$ for the BP/RP stars. Each of these results is approximately consistent with $1$, indicating an equal association between the GCs and the comparison and validation stars, as is expected given that these are in fact subsets of the same sample (see Section~\ref{subsec:2D_method}).

As shown in the left panels of Figures~\ref{fig:APOGEE_trends} and \ref{fig:BPRP_trends}, a distinct pattern emerges between probability excess and GC orbital energy. 
Overall, for GCs with energies $\lesssim-1.8\times10^5~\mathrm{km^2s^{-2}}$, the probability of association with 2G-type field stars is consistently higher than that for the comparison sample, marked by median probability excesses $>1$; for clusters with energies greater than this value, the probability excess drops below the validation line at $1$, marking a lower association between GCs and 2G-type field stars than the rest of the halo. The precise location of this boundary varies slightly between the APOGEE and BP/RP data, although there is a notable consistency in the results from the BP/RP data across the three clustering spaces used. 
In the BP/RP sample, the median excess association of the high-[N/O] field stars for the most bound GCs ($E\lesssim-2.5\times10^{-5}$) again drops below $\sim1$, which is likely because the BP/RP sample does not populate the very lowest energies, unlike the APOGEE stars (see Fig.~\ref{fig:ELz_hist}).
Overall, these results are consistent with the findings of previous works, particularly those that note an increasing fraction of N-rich stars towards the Galactic center \citep{Schiavon_2017,Horta_nrich_stars} and that link these GC-born stars to the \textit{in-situ} halo, \textit{Aurora} \citep{belokurov_aurora,belokurov_kravstov_nitrogen}. In agreement with these works, we interpret the finding that the lowest energy clusters have the strongest association with high-[N/O] stars in the halo as a signature that these stars are a component of \textit{Aurora} and do not contribute substantially to the accreted halo; we explore this interpretation further in Fig.~\ref{fig:insitu_accreted_4D}.

There is also a clear correlation between initial cluster mass (from the GC catalog, see Section~\ref{subsec:GC_data}) and probability excess, with the most initially massive clusters having the highest median probability excess for association with N-rich field giants. Below $M_\textrm{initial}\approx10^{6}~\mathrm{M}_\odot$ the median cluster probability for association with high-[N/O] field stars is lower than that for the comparison sample; above this boundary, the median probability excess increases to a value between $\sim1.5$ and $4$. Clearly, the high-[N/O] stars share similar integrals of motion with the most initially massive clusters and not with the smaller clusters. The relationship between probability excess and initial cluster mass is particularly notable given that the fraction of 2P (high-[N/O] stars) within clusters has been found to correlate with initial GC mass \citep{Gratton_what_is_GC}. Based on observational results that GCs lacking multiple populations have $M_\textrm{initial}\lesssim1-1.5\times10^5M_\odot$, it has been suggested that this initial mass represents a boundary lower than which multiple populations do not form \citep{Milone_2020, Lagioia_2024}. Interestingly, below this threshold of $M_\textrm{initial}=10^5M_\odot$, the probability excesses are close to 0 in both APOGEE and the BP/RP data, indicating almost no association with the N-rich field stars, and for the more initially massive GCs which are thought to have formed more 2P stars, the median probability excess is much higher. As we discuss later in the work, the difficulty of confidently claiming that a particular star was born in any single cluster makes it challenging to claim that this pattern is causal, but it is a nonetheless an interesting result that may corroborate theoretical predictions for the production of 2P stars. More investigation on this topic is, of course, necessary. We note that correspondingly strong trends for the most part do not exist with \emph{current} mass; the most massive GCs at present do not tend to have the highest associations with the high-[N/O] field stars, with the exception of a higher excess association between the very highest mass GCs and only the BP/RP---not the APOGEE---N-rich stars.

Finally, because the N-rich field giants are ``runaways'' from their GCs of origin and thus represent mass lost by the cluster, we show probability excess as a function of the change in cluster mass, $\Delta M=M_\textrm{current}-M_\textrm{final}$. Again, there is a clear pattern in which the GCs which have experienced the greatest mass loss have the strongest excess association with the high-[N/O] stars in the halo. In APOGEE, the median probability excess rises above $1$ for clusters which have lost $\gtrsim10^6~\mathrm{M}_\odot$, although the value is slightly lower for associations in the $E-L_z$ plane and slightly higher in the 4D $\vec{J}-\textrm{[Fe/H]}$ space. For the BP/RP stars, this transition to probability excesses $>1$ occurs closer to $\Delta M \gtrsim10^{5.7}~\mathrm{M}_\odot$. Across both the APOGEE and BP/RP data, these results indicate that the high-[N/O] stars share integrals of motion and metallicities with the clusters that have lost the most mass over the course of their lifetimes.

To further illustrate the connection between the high-[N/O] giants and the \textit{in-situ} GCs, we examine the clusters' probability excesses separately by their origin. Using the globular cluster classifications from \citet{Belokurov_2024_GCs}, we show in Fig.~\ref{fig:insitu_accreted_4D} the probability excess from the 4D $\vec{J}-\textrm{[Fe/H]}$ space per cluster for \textit{in-situ} and accreted separately. The distributions of excesses clearly differ between the two, with most of the accreted GCs having close to zero probability excess relative to the comparison set and the \textit{in-situ} clusters having a split between very low probability excesses and $\epsilon_{c,~\textrm{excess}}>1$. That these trends existed when the associations were made in the $E-L_z$ plane is unsurprising, as the classification between accreted and \textit{in-situ} GCs is itself made in this space and high-[N/O] field stars have already been noted to mostly populate the inner galaxy \citep{Martell_2010,Schiavon_2017,Horta_nrich_stars} and likely to belong to \textit{Aurora} \citep{belokurov_kravstov_nitrogen}. That this relationship between the \textit{in-situ} GCs and N-rich field stars persists into this chemodynamical space further highlights the connection between these GC-origin stars and the \textit{in-situ} halo--and the lack of such stars in the accreted halo.

Finally, regarding tagging individual N-rich stars in the halo to a specific cluster in which they were born, the close associations between these stars and the population of \textit{in-situ} clusters as a whole proves a challenge. In Figures~\ref{fig:4D_prod_APOGEE}~and~\ref{fig:4D_prod_BPRP}, we show the integrated products of each individual star's and GC's PDFs ($\epsilon_{c,~s}$) from the $\Vec{J}$-[Fe/H] space for the APOGEE and BP/RP samples, respectively. The results are sorted by both star and GC orbital energy and clearly illustrate that for any given star, there typically are associations with multiple GCs at a similar energy. This holds especially true at the lowest energies. These integrated products $\epsilon_{c,~s}$ are sufficiently similar for so many GCs that it would be difficult to securely claim that a single star originated within any particular GC. We interpret this result as an indication that the many of the \textit{in-situ} clusters and high-[N/O] field stars occupy a sufficiently small chemo-dynamical space and have sufficiently large errors that individual birth associations become difficult to confidently assert; this challenge becomes even more substantial for the BP/RP stars, which have much more poorly constrained metallicities than those from APOGEE. We thus restrict this study to the bulk associations between the GC-origin stars in the halo and the surviving Galactic globular clusters. Although we do not rule out the possibility that some of the stars in this work do originate in a GC to which we tag them, as we outline in Sections~\ref{subsec:df} and \ref{subsec:galactic_bar}, there is good reason to be cautious about these specific associations.

As is also evident in Figures~\ref{fig:4D_prod_APOGEE}~and~\ref{fig:4D_prod_BPRP}, although the lowest energy stars have strong associations ($\epsilon_{c,~s}$) with multiple clusters, the higher energy N-rich field stars have relatively fewer high values of $\epsilon_{c,~s}$ with any given cluster. It is again impossible to rule out that these stars originated in surviving clusters, for their lack of shared properties with the Galactic GCs could be a result of the integrals of motion of the clusters and N-rich stars not having been conserved since the escape of the star from its birth cluster (as we discuss further in Section~\ref{sec:discussion}). Nonetheless, these ``unassociated" stars with such strong chemical signatures of having been born in a GC could be the remnants of many clusters that once existed in the Galactic halo and have since dissolved entirely, as has been suggested by multiple previous works \citep{Schiavon_2017, belokurov_kravstov_nitrogen} and was a suggested possible conclusion for similar findings from other studies that looked at associations between N-rich stars and GCs \citep{Savino_Posti_2019,Hanke2020_method,Xu2024_method}. In fact, even some of the stars ``tagged" to many clusters could still have originated in dissolved GCs given how crowded the actions-space is at low energies (see Fig.~\ref{fig:actions_hist}). More work is yet needed to confirm this result, but there is mounting evidence that there were once many more globular clusters in the Milky Way.

\section{Discussion}
\label{sec:discussion}

\subsection{The GCs that Formed the High-[N/O] Field Stars}
\label{subsec:massive_GCs}

\begin{figure}
	\includegraphics[width=0.7\columnwidth, alt={As compared to the original probability excesses calculated with the full comparison samples in Figures 5 and 6, the trends in median probability excess here are clearly different. In particular, although the median excesses are still <1 at low energies, they only increase slightly at or slightly before the Solar energy. For  the APOGEE stars, the median probability excess never reaches the equal association line at 1 even for the lowest energy clusters. For the BP/RP stars, the median probability excess appears approximately consistent with 1 at energies lower than Solar, with some variation. These results are approximately consistent across the three clustering spaces. Overall, it is clear that the median probability excess never reaches as high values as those in Figures 5 and 6.}]{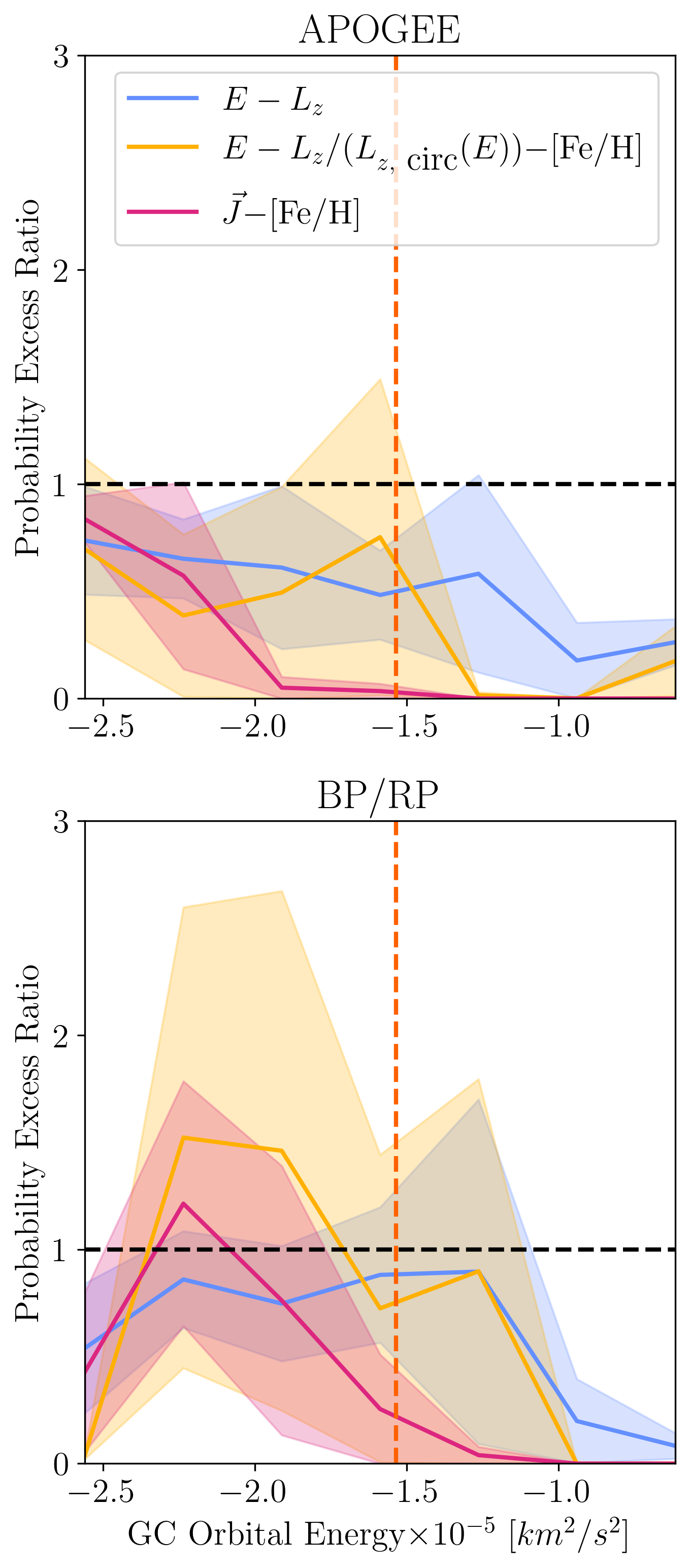}
    \centering
    \caption{As in the left panels of Figures~\ref{fig:4D_prod_APOGEE} and \ref{fig:4D_prod_BPRP}, this shows the probability excess per GC as a function of GC orbital energy for associations in the $E-L_z$ plane (blue), the 3D $E-L_z/L_{z,~\textrm{circ}}(E)-\textrm{[Fe/H]}$ space (yellow), and the 4D action-metallicity space (pink), now with a comparison sample that is selected to have the same energy distribution as the high-[N/O] field giants. The probability excesses are shown separately for the APOGEE and BP/RP samples in the upper and lower panels, respectively. As before, the solid lines indicate the median probability excess, and the shading marks the median absolute deviation. The vertical orange dashed line marks the Solar energy.}
    \label{fig:4D_trends_energy_matched}
\end{figure}

The connections between low energy, initially massive GCs and the high-[N/O] stars in the halo, which persist across both energy and angular momentum and the chemo-dynamical action-metallicity space, point to these massive clusters having been a significant contributor to the population of these 2P stars that are now observed in the halo. It is possible that some significant fraction of these stars were also formed in now entirely disrupted clusters that once occupied approximately the same chemo-dynamical space that is still filled by the surviving GCs. These clusters would have been slightly too small to survive the busy neighborhood of the inner Galaxy but were still initially massive enough to produce a second generation. Especially in the very innermost halo, even relatively massive ($M_\textrm{initial}\approx10^5-10^6~\mathrm{M}_\odot$) clusters could theoretically have been entirely disrupted \citep[see Fig.~7 of][]{Baumgardt_2019}.

In Section~\ref{sec:results}, we highlight that probability excess for association between GCs and N-rich field stars increases for the lowest energy GCs, the most initially massive GCs, and the GCs which have lost the most mass. Here, we suggest that these trends are all in fact interconnected: the most bound clusters tend to have high initial masses, as any GCs smaller than $M_{\mathrm{initial}} \approx 10^6~\mathrm{M}_\odot$ would have been dissolved entirely, leaving only the most initially massive as the survivors. These survivors have themselves lost a great deal of their mass. Thus, given that there is an increase in probability excess for the lowest energy clusters, it should be unsurprising that the excess is also highest for the most initially massive GCs and those which have lost the most mass, as these are the properties of the lowest energy clusters.

In light of this result, we return to the question as to why the N-rich second generation stars dominate the inner (\textit{in-situ}) halo, given that there are clearly many accreted GCs, many of which were sufficiently massive to produce nitrogen-enhanced second generation stars. The results of this work could corroborate the explanation proposed by \citet{belokurov_kravstov_nitrogen}, wherein the vast majority of high-[N/O], GC-origin stars originated in \textit{Aurora} because only the low energy GCs experienced sufficiently strong tidal forces to lose substantial numbers of their N-rich stars, especially given that second generation stars tend to be centrally concentrated in some GCs \citep[e.g.,][]{Leitinger_gc_pop_distribution,Mehta_2025}. By contrast, the Gaia Sausage/Enceladus \citep[GS/E][]{Belokurov_GSE,Helmi_GSE}, although it has formed massive clusters capable of producing substantial numbers of second generation stars, never stripped its clusters efficiently of these stars, either before or after its accretion by the Milky Way \citep[as is evident by the lower mass losses experienced by higher energy clusters in][]{Baumgardt_2019}. For GCs with less than Solar energy, the mean mass lost was approximately $82\%$, with the ``survivors" being the most initially massive clusters. By contrast, the higher energy GCs ($E>E_\odot$, many of which were accreted) lost an average $67\%$ of their initial mass, but many of these were smaller clusters to begin with as compared in the inner MW GCs. The GS/E clusters that \emph{were} destroyed entirely were likely the smallest ones which had the fewest second generation stars, leaving little chemical trace of their existence in the accreted halo. The same is likewise probably true for other, smaller accreted structures.

However, it has been established by previous works that high-[N/O] field stars have a different energy distribution from the rest of the halo as a whole; this is true both for the APOGEE stars \citep[see][]{belokurov_kravstov_nitrogen} and for the BP/RP candidates \citep{Kane_2024}. Given that these N-rich stars are more concentrated at low energies than the comparison sample and that our probability excesses for association with GCs rise notably for the lowest energy GCs, it is worth examining in some more detail whether this trend is driven solely by the energy distribution of the high-[N/O] field stars rather than by specific agreement between the metallicities and integrals of motion of these stars and the GCs. To this effect, we construct secondary comparison samples, again separately for APOGEE and the BP/RP data, such that it has approximately the same energy distribution as the high-[N/O] stars. These secondary comparison samples are selected as a subset of the full comparison sample detailed in Section~\ref{subsec:data} and are comprised of 1\,189 stars for APOGEE and 19\,078 stars for the BP/RP sample.

\begin{figure*}
	\includegraphics[width=1.85\columnwidth, alt={Below Solar energy, GCs' energies and the z-component of their angular momenta change over the span of the orbit integrations, with the extent of those changes increasing with decreasing energies. The orbital energies almost universally decrease as a result of dynamical friction, with the lowest energy clusters losing approximately 6000-8000 km^2/s^2 of their initial energy. Lz values of the low energy GCs can both increase and decrease, with the changes being on the order of 400 kpc km/s or less. For GCs with current energies higher than Solar, the changes in both E and Lz are much less pronounced. The low energy clusters have almost universally the highest initial masses (above 1,000,000 Solar masses), while lower energy clusters exhibit a broader range in initial masses (down to 10,000 Solar masses or below).}]{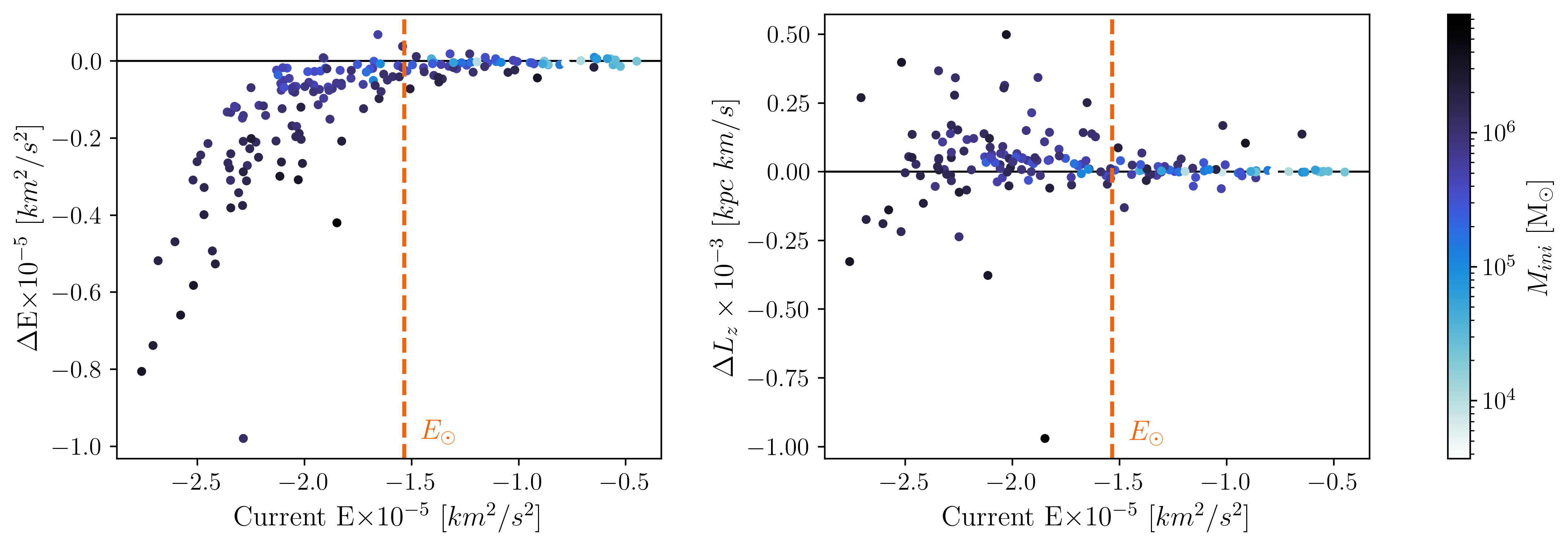}
    \caption{The changes in GC orbits over their lifetimes as produced via the orbital integrations with dynamical friction detailed in Sec~\ref{subsec:df}. The two panels show the change in each cluster's orbital energy (left) and angular momentum, $L_z$ or $J_\phi$, (right) versus the current energy of the cluster. The cluster orbital properties are re-calculated with Astropy's Galactocentric coordinates and in the \citet{McMillan2017} potential for consistency and easy comparison to the results in previous sections. In both panels, the Solar energy is marked with the vertical red, dashed line. Color-coding represents the log-scaled initial mass of the cluster.}
    \label{fig:gc_orbit_df}
\end{figure*}

Using this new comparison sample, we perform the same procedure outlined in Section~\ref{sec:data_and_methods} to calculate a revised probability excess per GC in each of the three clustering spaces. We show the results of this test separately for the APOGEE and BP/RP stars in Fig.~\ref{fig:4D_trends_energy_matched}. As is evident from the figure, the strength of the increase of median probability excess in low energy GCs is notably decreased; the low excesses $<<1$ for the high energy clusters remains with this secondary comparison sample, but the probability excess for GCs with $E\lesssim-1.8\times10^5~\mathrm{km^2s^{-2}}$ is only $\sim1$ for the BP/RP stars and slightly lower still for the APOGEE sample. These results indicate that the high-[N/O] field stars do not have any substantially greater association with the surviving inner Galaxy clusters than a selection of chemically typical field stars that share a similar energy distribution; in other words, the high-[N/O] stars share no more dynamical similarity to the surviving GCs than a selection of typical inner Galaxy, \textit{Aurora} stars, in spite of the chemistry indicating that they ought to have originated in GCs. Some of this effect may be an interesting confirmation that even among chemically typical stars in the \textit{Aurora}, many were formed in GCs, with \citet{belokurov_kravstov_nitrogen} estimating that fraction to be $\geq50\%$. Even so, it is still reasonable to expect a \emph{higher} fraction---likely all---of the high-[N/O] stars to be formed in GCs, leading to the expectation that some probability excess should persist even when we select an energy-matched comparison sample. That this is not the case indicates that the associations between the massive, inner Galaxy clusters and the N-rich field giants that we explore in Section~\ref{sec:results} are primarily driven by the differing energy distributions between the high-[N/O] stars and the comparison sample when that comparison sample is comprised of stars from across the halo, not just \textit{Aurora}. Nonetheless, this energy distribution of the high-[N/O] stars itself, as evidenced by the excess association of these stars with low energy GCs as compared to the halo stars as a whole in Section~\ref{sec:results}, is likely still indicative that in bulk these stars formed in the inner Galaxy GCs.

That the high-[N/O] stars are no more associated with the surviving GCs than other \textit{Aurora} stars could reasonably mean one of three things. First, some extremely high fraction of \textit{Aurora}, of which our new energy-matched comparison sample is likely predominantly composed, was formed in GCs, and we are uncovering the signature of this effect in the chemically typical stars. Although we propose that this is a contributing factor to explain the decrease in probability excess with the energy-matched comparison sample, it seems improbable that \emph{such} a high fraction of the [N/O]-typical stars formed in the surviving GCs, as there are expected to be many entirely destroyed GCs from the inner Galaxy, and many nitrogen-typical stars should also be expected to form in the field. Second, the high-[N/O] stars could have originated themselves from entirely disrupted GCs, but given that the surviving inner MW GCs are thought to have lost a large fraction of their mass and are observed to host large numbers of 2P stars, it seems reasonable to expect that at least some substantial number of the high-[N/O] stars in the halo did in fact form in these surviving GCs. Finally, the most likely reason, which we explore in more detail in the following two sections, is that the integrals of motion of stars and GCs are not truly conserved in the inner MW, meaning that associations between a high-[N/O] star and its GC of origin is eventually reduced or ``erased" after it escapes from the cluster.

\subsection{The Effect of Dynamical Friction on GC Orbits}
\label{subsec:df}

Throughout this work thus far, we have implicitly assumed that the GCs and their escapees have actually conserved their integrals of motion since the time the stars escaped. As was briefly mentioned in Section~\ref{sec:data_and_methods} and has been discussed in other similar works \citep[e.g., see][]{Savino_Posti_2019}, this is not necessarily the case. Here, we further examine the impact of dynamical friction on the conservation of the properties of the GCs' orbits; in the next section, we also discuss the effect of the bar on ``conserved" integrals of motion. The results of this section are intended to demonstrate the scope of the effect that dynamical friction can have on some GCs’ orbits over the span of their lifetimes. As such, because we are \emph{not} attempting to actually retrace the original orbits of these GCs, we make several simplifying assumptions in order to produce an illustrative calculation for the purposes of this discussion.

The following procedure was adopted: from the GCs’ current positions and velocities, their orbits were backwards integrated in the potential from \citet{Irrgang_2013}. Chandrasekhar dynamical friction is included via the prescription from Equation 7-18 of \citet{Binney_Tremaine_1987}, which is:
\begin{equation}
    \frac{d\vec{v}_\mathrm{GC}}{dt}=-\frac{4\pi \ln\Lambda G^2 \rho M_\mathrm{GC}}{v_\mathrm{GC}^3}(\erf{X}-\frac{2X}{\sqrt{\pi}}e^{-X^2})\vec{v}_\mathrm{GC}
\end{equation}
where $M_{GC}$ is the current mass of the GC and $\vec{v}_{GC}$ is the velocity of the GC. The background density $\rho$ is taken as the local density at the position of each cluster at any given timestep as given by the \citet{Irrgang_2013} potential. Following from \citet{Binney_Tremaine_1987}, the Coulomb logarithm is taken as $\ln\Lambda\approx10$ for the GCs. The parameter $X$ is defined as $v_\mathrm{GC}/(\sqrt{2}\sigma)$, where $\sigma$ is the velocity dispersion of the background particles (i.e., field stars) which we take as the circular velocity of the Milky Way at the position of the GC.

Mass loss in the clusters, both from stellar evolution and dynamical mass loss (i.e., the loss of individual stars due to two-body relaxation and the interaction with the tidal field of the Milky Way), is also included. 
Using PARSEC isochrones \citep{Bressan_2012_PARSEC}, we first estimate the fractional mass loss rate as a function of time of an isolated population of stars due to stellar evolution.
Then, according to \citet{Baumgardt_Makino_2003}, we calculate the dissolution time of the cluster given the current orbit, which also yields a current dynamical mass loss rate for the GC, $\tfrac{dM_\textrm{dyn}}{dt}$. The mass of each GC is then increased at each timestep by a value of $\Delta M=M_\textrm{current}+\Delta M_\textrm{SEv}(t)+\Delta M_\textrm{dyn}$, where $M_\textrm{current}$ is the mass of the GC at the start of each timestep, $\Delta M_\textrm{SEv}(t)$ is the fractional mass lost at each timestep due to stellar evolution, and $\Delta M_\textrm{dyn}$ is the dynamical mass lost during the timestep given the GC's orbit. We backwards integrate the GC's orbit, adding mass accordingly, until the individual birth time for each cluster is reached.

We show the results of this experiment in Figure~\ref{fig:gc_orbit_df}. As can be seen in the left panel of the figure, the orbital energy changes over the lifetime of the cluster are the most prominent for GCs with current energies $\lesssim-2\times10^{-5}~\mathrm{km^2/s^2}$; by comparison, clusters with energies $\gtrsim-1.5\times10^{-5}~\mathrm{km^2/s^2}$ are much less impacted by the effects of dynamical friction, although the more initially massive clusters are more affected than the smaller clusters. For some clusters, the impact of dynamical friction on their orbital energies is extreme, with 32 GCs experiencing a decrease in energy of $10\%$ or more of their current orbital energy. This effect can also be observed in other integrals of motion; in the right panel of Fig.~\ref{fig:gc_orbit_df}, the GCs' $L_z$ are clearly also not conserved, with the effect again being most pronounced for the low energy and initially massive GCs---the very clusters to which we associate the high-[N/O] field stars.

\begin{figure*}
	\includegraphics[width=2\columnwidth, alt={Three panels show the probability density functions (PDFs) of the GCs, the APOGEE N-rich field stars, and BP/RP N-rich field stars in the space of [Fe/H] versus the Jacobi integral. The PDFs of of the GCs for the most part appear as small oval clouds with widths of approximately 0.25 dex in metallicity and relatively good constraints on their Jacobi integral, although a few of the PDFs appear "fluffier" and less constrained. The GCs as a population span the full range of metallicities shown, from -0.9 dex to less than -2.25, and also a range of Jacobi integral values. The PDFs of the APOGEE N-rich stars for the most part appear to occupy only a single 0.05 dex bin in metallicity each, but the energy spread of each star is much broader. The metallicities clearly do not extend beyond approximately [Fe/H]=-2. The PDFs of the BP/RP stars in this space are clearly very poorly constrained in the metallicity dimension and span a wide range of [Fe/H] values; this effect makes it difficult to distinguish individual PDFs in the visualization, as they overlap a great deal. As in the energy spaces in Figures 1 and 3, the BP/RP stars do not reach as low values of the Jacobi integral as do the APOGEE N-rich stars, but the metallicity errors extend below [Fe/H]=-2.}]{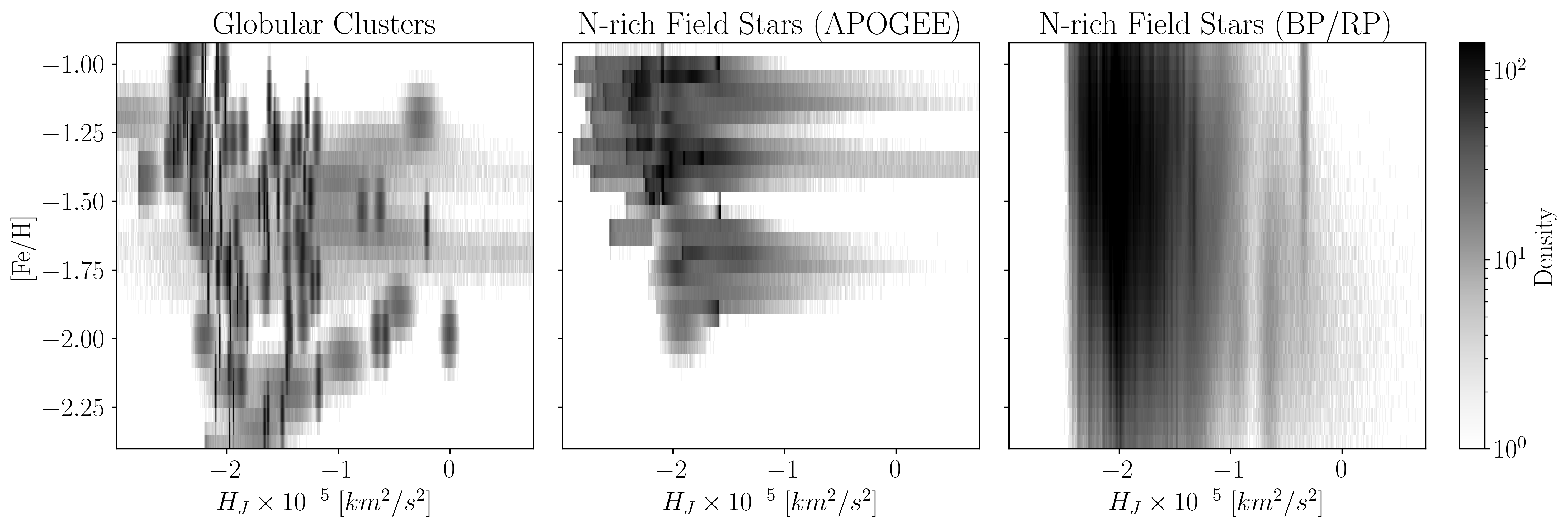}
    \caption{As in Section~\ref{subsec:2D_method}, the 2D histograms of the distributions of the GCs (left), the high-[N/O] field stars from APOGEE (middle), and the high-[N/O] stars from the BP/RP sample (right) in the [Fe/H]-Jacobi energy space. Note the much larger spread of the BP/RP stars in metallicity, corresponding to the greater uncertainties relative to the GCs or the APOGEE stars. Some of the highest $H_\mathrm{J}$ bins are excluded for visualization purposes.}
    \label{fig:Ej_example}
\end{figure*}

\begin{figure*}
	\includegraphics[width=7in, alt={It is immediately apparent that the trends in median probability excess are very similar when shown relative to either orbital energy or Jacobi integral.  Below Solar energy or Solar Jacobi integral, the median probability excesses for APOGEE or the BP/RP stars are between 0.25 and 0.75; at just slightly above Solar energy, the median excesses increase above the validation line at 1 to approximately 2 to 2.5. For both orbital energy and Jacobi integral, the median probability excess for GCs with the BP/RP N-rich stars drops below 1 again at the very highest values; this behavior is not reflected in the probability excesses calculated with the APOGEE stars. At the very highest orbital energies, the median probability excess for APOGEE and the BP/RP daya increases to approximately 1.25, but this does not occur at the highest values of the Jacobi integral. Relative to initial mass, the median trends for both APOGEE and BP/RP remain relatively constant at probability excess of approximately 1.5, with median absolute deviations of approximately 1. The BP/RP probability excess drops to approximately 1 for the lowest initial masses (approximately 10,000 Solar masses) and the highest (greater than 1,000,000 Solar masses).}]{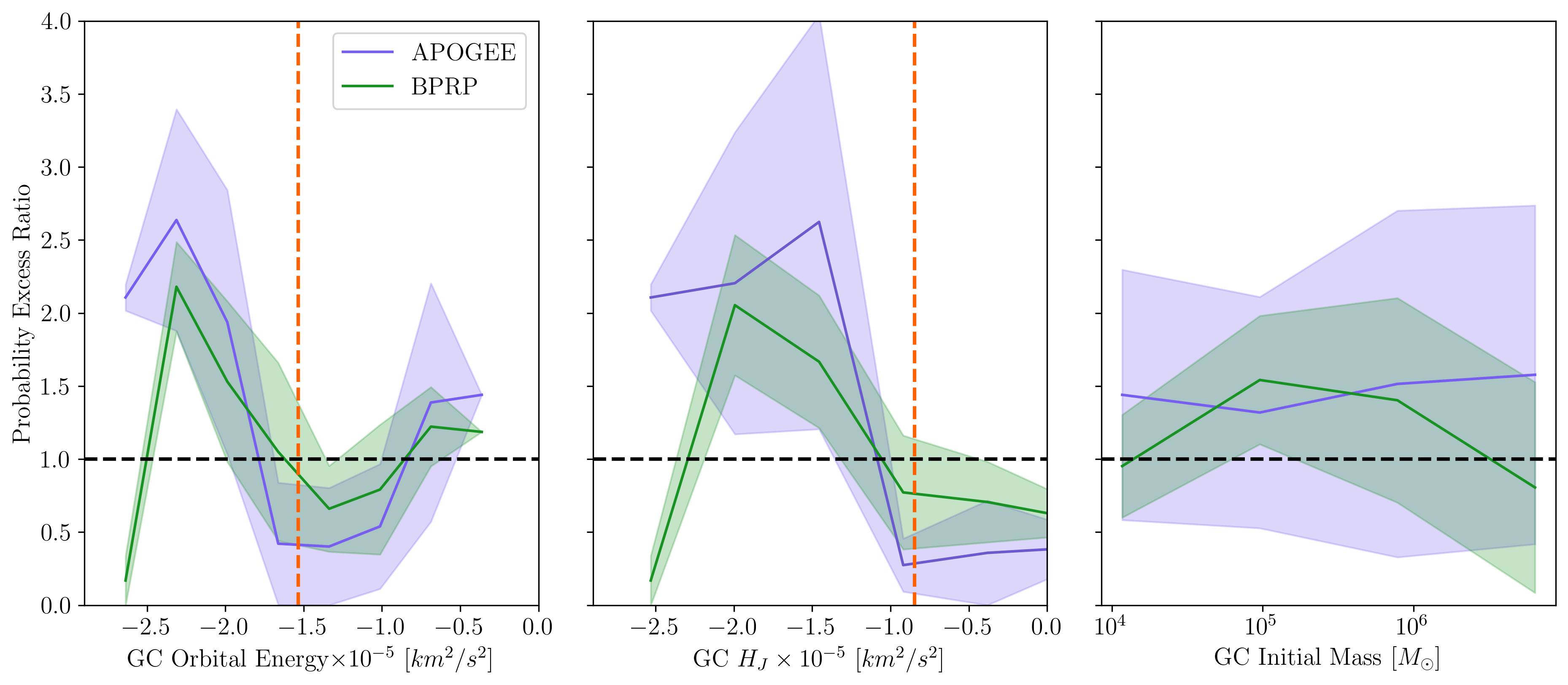}
    \caption{The results of defining associations via $H_\mathrm{J}$ and [Fe/H], these are the probability excess per GC as a function of energy $E$ (left), $H_\mathrm{J}$ (middle), and GC initial mass (right). The probability excess per cluster with the APOGEE field stars is shown with the blue circular markers and with the BP/RP field stars with the green triangles. The lines again indicate the median trends, with the shading showing the median absolute deviation in each bin. The orange dashed vertical lines in the middle and right panels denote the Solar Jacobi integral and orbital energy, respectively.}
    \label{fig:Ej_trends}
\end{figure*}

The substantial changes in the orbits of the Galactic globular clusters provide another reason to be cautious of even the 4D chemodynamical associations described in Section~\ref{sec:results}, especially given that these changes due to dynamical friction are most prominent for the massive, inner Galaxy GCs that we tag as the origin of many of the high-[N/O] stars in the field. There are several potential interpretations of these results. First, it is possible that some number of these second population field stars escaped from their birth clusters relatively recently, meaning that that particular GC’s and star’s integrals of motion remain relatively similar and that the associations we uncover here are genuine. Second, one could assume the associations we find in this work to be the \emph{minimal} present-day associations between a GC and its escapees, and if the orbital history and escape time of the star could be accurately reconstructed, the probabilities of association would be much higher. Finally, it remains possible that these stars have in fact been erroneously been tagged to their clusters, particularly at the inner Galaxy, and that for any given high-[N/O] star, the true association is actually with a GC at a lower energy than the ones to which we tag it here. We leave this as a possibility but assert that it does not change the primary conclusion of this work: namely, that massive, inner Galaxy clusters are the primary formation site of many of the second generation GC stars in the halo.

\subsection{The Galactic Bar and Non-conservation of Integrals of Motion}
\label{subsec:galactic_bar}

Throughout this work, we have assumed the static, axisymmetric \citet{McMillan2017} Milky Way potential for the calculation of our integrals of motion.
However, recently, \citet{Woudenberg_2025_bar} and \citet{Dillamore_2025_bar} have both highlighted the influence of the bar as another source of change in integrals of motion, this time affecting both GCs and field stars. Both of these works find that the bar can disperse substructure in the space of integrals of motion and that this effect is the most pronounced for the lowest energy stars. Inconveniently, these lower energies are where most of the high-[N/O] field stars in this work reside \citep[][see also Fig.~\ref{fig:ELz_hist}]{belokurov_kravstov_nitrogen, Kane_2024}, suggesting that the bar may have a pronounced effect here. In fact, \citet{Dillamore_2025_bar} specifically address the fact that high-[N/O] field stars are unlikely to continue to share integrals of motion with their GC of origin due to the effects of the Galactic bar.

However, \citet{Dillamore_2025_bar} note that while the bar causes substructure to disperse in, for instance, the $E-L_z$ plane, the Jacobi integral remains constant in the case of a bar with a constant pattern speed. Even in the presence of a slowing Galactic bar \citep[see e.g.,][]{Zhang_2025_radial_migration}, the Jacobi integral is far less affected than other integrals of motion. \citet{Dillamore_2025_bar} thus suggest that identifying substructure with the Jacobi integral and chemical information may be less affected by the influence of the bar. 

Following these results, we now apply this suggestion to observational data by employing our clustering algorithm from Section~\ref{subsec:2D_method}, now adapted to a 2D Jacobi integral-[Fe/H] plane. We take the same steps of generating 10\,000 samples from the astrometry and metallicity for each object, in this case using the astrometry to calculate the Jacobi integral, $H_\mathrm{J}=E-\Omega_\mathrm{b}L_z$, in the \citet{McMillan2017} potential. Again following from \citet{Dillamore_2025_bar}, we adopt a current bar pattern speed $\Omega_\mathrm{b}$ of $34.5~\mathrm{km/s/kpc}$, which is also approximately consistent with recent measurements of the Galactic bar \citep{Binney_2020,Zhang_2024_bar,Dillamore_2025_dynamical_streams,Chiba_2021}. The samples are then assigned into 500 bins in the $H_\mathrm{J}$ dimension ranging from the 1st to the 99th percentile of GC sample values and into 30 bins in the [Fe/H] dimension, again ranging from the 1st to 99th percentile of GC samples. The resulting $H_\mathrm{J}-$[Fe/H] plane of the associated samples from the GCs and the high-[N/O] field stars, both from APOGEE and the BP/RP data, is shown in Fig.~\ref{fig:Ej_example}; the figure also clearly illustrates how much smaller the uncertainties are for the metallicities of the APOGEE stars and GCs as compared to the BP/RP stars. The calculation of association between field stars and GCs and the corresponding probability excess of association with the high-[N/O] field stars per cluster then follows the same procedure as that outlined in Sec.~\ref{subsec:2D_method}.

The results of this test, again presented as the probability excess of association between the N-rich field stars and the GCs as a function of the GC properties, are shown in Fig.~\ref{fig:Ej_trends}. There is a remarkable consistency in the trends of probability excess with orbital energy from the $H_\mathrm{J}-\textrm{[Fe/H]}$ space as compared to some of the other spaces outlined in Section~\ref{sec:results}, with the median probability excess increasing sharply for clusters with orbital energies slightly below Solar ($\lesssim1.8~\mathrm{km^2/s^2}$. Interestingly, as is shown in the middle panel of Fig.~\ref{fig:Ej_trends}, a very similar trend of the probability excess exists with cluster Jacobi integral, with the median probability excess again increasing for clusters with $H_\mathrm{J}\lesssim-1.0~\mathrm{km^2/s^2}$, slightly lower than the Solar Jacobi energy. Both of these patterns exist for both the APOGEE and BP/RP stars, although the probability excess drops below the validation line of $1$ again at the lowest values of $E$ and $H_\mathrm{J}$ for the BP/RP sample, again likely reflecting the fact that the BP/RP stars do not populate as low energies as the APOGEE stars (see Fig.~\ref{fig:ELz_hist} and further discussion above in Sec~\ref{sec:results}). By comparison, the trends of probability excess with GC initial mass are not nearly as pronounced here, suggesting that energy is the main driver for association; this is likely unsurprising as $H_\mathrm{J}$ itself is used to calculate associations.

Given the effects of both dynamical friction and the Galactic bar, it seems unlikely that integrals of motion are truly conserved, especially at the inner Galaxy where both of these effects are the most pronounced. Therefore, the underlying assumption of this work---that “runaway” high-[N/O] stars ought to share integrals of motion with their birth cluster---is likely not correct unless the star escaped relatively recently. We thus encourage caution when it comes to any clustering of GC debris in integrals of motion spaces, and reiterate that it is likely difficult---or impossible---to tag a specific star to a specific cluster in the context of this work. Nonetheless, the larger-scale conclusion of this work, namely that the low energy, inner Galaxy GCs were likely the main source of the high-[N/O] stars in the field, is probably more robust against these effects as it has more to do with the general location of the N-rich stars in the integrals of motion spaces than tagging to any specific cluster. This assertion is supported by the persistence of these results in the Jacobi integral-[Fe/H] space, which should be less affected by the influence of the Galactic bar, though the Jacobi integrals of the GCs may nonetheless not be conserved due to dynamical friction. 

In some respect, that the uncertainties are large on the stars' and clusters' integrals of motion is useful here, as it means that even if a GC's orbit has changed, it could nonetheless overlap with its escapee stars. However, this limits the utility of tighter constraints on the astrometric and distance uncertainties from future \textit{Gaia} data releases, as even if stronger associations with some clusters can be found, these associations may not be genuine. Some effort could be made to retrace the orbits of the clusters under the influence of dynamical friction, as has been done earlier in this work, and clustering with these backwards-integrated orbits using the Jacobi integral may be more robust. However, maintaining accuracy over long timescales of backwards integration will demand a number of constraints: on the evolution of the bar pattern speed, on the changing MW potential, and so forth. Another interesting prospect for the future may include using neutron-capture element abundances ($r-$ or $s-$process) as an additional chemical tag alongside Fe, given that most GCs exhibit only small spreads in these heavy elements \citep[e.g.,][]{Monty2024,SchiappacasseUlloa_HeavyElements}.

\section{Conclusions}
\label{sec:conclusions}

In this work, we associate populations of stars with anomalous chemistry consistent with the second generation of stars in globular clusters, to intact Milky GCs in chemo-dynamical space. The method relies upon similarities in the clusters' and stars' energy and $L_z$ values (Section~\ref{subsec:2D_method}), their energies, the scaled $z$-component of their angular momenta, and metallicities, and their actions and metallicities (Section~\ref{subsec:3D4D_method}) based on Monte Carlo sampling from the errors on the observables of each object. The associations between high-[N/O] field stars and GCs are then scaled by the associations between a chemically typical comparison sample of giants in the halo to determine a \emph{probability excess} per cluster, a quantity that approximates the degree to which any given cluster shares a stronger association with the N-rich stars than with the halo overall. This method allows us to examine the contribution of each globular cluster to the Milky Way halo.

We summarize our findings as follows:

\begin{enumerate}
    \item We note a distinctive correlation between the integrals of motion and metallicities of the nitrogen-rich field stars and surviving globular clusters with certain properties. In particular, we note that the clusters with the lowest orbital energies and highest initial masses have the highest “probability excess” of association with the N-rich field stars as compared to the halo as a whole. These trends are largely consistent across the three clustering spaces we employ and between the APOGEE and BP/RP stars. This correlation between the low energy clusters and theorized GC-born stars in the field is consistent with previous findings that associate most N-rich field stars with \textit{Aurora}, the in-situ MW halo and, more broadly, with the fact that N-rich stars are more frequent closer to the Galactic center \citep{belokurov_kravstov_nitrogen,Horta_nrich_stars,Schiavon_2017}. These trends are such that almost all accreted globular clusters have excess association with the high-[N/O] field stars $<1$, whereas the \textit{in-situ} GCs have a substantial number of clusters with an excess probability $>1$. The difference in association between the N-rich field stars and the accreted versus \textit{in-situ} clusters highlights the contribution of GCs to \textit{Aurora} and the relative lack of such a contribution in accreted systems.
    \item Even in the more complex action-metallicity space, it is difficult to confidently tag high-[N/O] field giants as having originated in any specific Galactic GC. This challenge arises because the uncertainties in the stars' and GCs' properties combined with the similarities in many of their orbits and metallicities means that one star can have a PDF $P(\Vec{J},\textrm{[Fe/H]})$ that overlaps with many clusters. We cannot exclude the possibility that some of these stars originated in a cluster with which they are associated via this method but merely advise caution in doing so with confidence.
    \item We suggest that the high-[N/O] field stars are more associated with the inner Galaxy GCs as compared to the comparison sample of other halo stars because it is only in the central Milky Way that massive clusters, which produce many second generation, nitrogen-enhanced stars, are substantially disrupted, as was the hypothesis of \citet{belokurov_kravstov_nitrogen}.
    \item In Section~\ref{subsec:massive_GCs}, we illustrate that the increase in the excess association of the N-rich field stars with GCs for the low energy, most initially massive clusters does not persist if the comparison samples are selected to match the energy distribution of the high-[N/O] field stars. In essence, this illustrates that these trends in probability excess with GC properties (energy, initial mass, etc.) are themselves driven by the energy distribution of the N-rich field stars, which are more centrally concentrated in the Galaxy \citep{belokurov_kravstov_nitrogen, Kane_2024}, and that these GC-origin stars are hardly more dynamically associated with the clusters than their chemically-typical counterparts in the inner halo. The energy distribution of the N-rich field stars is such that it is nonetheless likely that inner Galaxy GCs were the predominant source of these stars \emph{en masse}, but disentangling more precise origins for these stars is difficult.
    \item One plausible explanation for the difficulty in finding excess association of the N-rich field stars with GCs as compared to other inner halo stars is that the GC ``runaways'' no longer share integrals of motion with their birth cluster. As we discuss in Sections~\ref{subsec:df}, dynamical friction is likely to have substantially affected the orbits of GCs, especially those of the inner Galaxy clusters which we think are the most likely origin of the N-rich field stars. The effect of dynamical friction is such that GCs gradually move further into the Galactic potential, with some 32 GCs having had initial energies $\geq10\%$ higher than their current values, and other integrals of motion can be similarly affected (see. Fig.~\ref{fig:gc_orbit_df}). GCs---especially inner Galaxy GCs--thus may no longer have the same integrals of motion that they once had when high-[N/O] stars escaped the cluster. Furthermore, as has been discussed in recent literature \citep{Woudenberg_2025_bar, Dillamore_2025_bar}, the influence of the Galactic bar means that integrals of motion for both stars and GCs are unlikely to be truly conserved, again with the most pronounced effect being present at the inner Galaxy. Per the suggestion of \citet{Dillamore_2025_bar}, we look for associations between GCs and N-rich field stars with [Fe/H] and the Jacobi integral, $H_\mathrm{J}$, which should remain less affected by the bar. In Fig~\ref{fig:Ej_trends}, we find that the similar trends of the most initially massive, inner Galaxy clusters have the highest excess association with the high-[N/O] field stars also exists in this space, suggesting that this bulk association is persistent against the affects of the Galactic bar. Nonetheless, both dynamical friction and the Galactic bar make precise tagging between GCs and their runaways a challenging---if not impossible---prospect.
\end{enumerate}

\section*{Acknowledgments}

We thank Adam Dillamore and Robyn Sanderson for useful discussions regarding the content of this paper. SGK acknowledges PhD funding from the Marshall Scholarship, supported by the UK government and Trinity College, Cambridge.
HZ thanks the Science and Technology Facilities Council (STFC) for a PhD studentship.

This work made extensive use of the Python packages \texttt{Numpy} \citep{harris2020array}, \texttt{Scipy} \citep{2020SciPy}, \texttt{Matplotlib} \citep{Hunter:2007}, and \texttt{Scikit-learn} \citep{scikit-learn}. This work made use of \texttt{Astropy}:\footnote{http://www.astropy.org} a community-developed core Python package and an ecosystem of tools and resources for astronomy \citep{astropy:2013, astropy:2018, astropy:2022}. This paper made used of the Whole Sky Database (wsdb) created by Sergey Koposov and maintained at the Institute of Astronomy, Cambridge with financial support from the Science \& Technology Facilities Council (STFC) and the European Research Council (ERC).

This work has made use of data from the European Space Agency (ESA) mission
{\it Gaia} (\url{https://www.cosmos.esa.int/gaia}), processed by the {\it Gaia}
Data Processing and Analysis Consortium (DPAC,
\url{https://www.cosmos.esa.int/web/gaia/dpac/consortium}). Funding for the DPAC
has been provided by national institutions, in particular the institutions
participating in the {\it Gaia} Multilateral Agreement. 

This work made use of data from the Apache Point Observatory Galactic Evolution Experiment \citep[APOGEE][]{APOGEE_DR17}. Funding for the Sloan Digital Sky 
Survey IV has been provided by the 
Alfred P. Sloan Foundation, the U.S. 
Department of Energy Office of 
Science, and the Participating 
Institutions. 

SDSS-IV acknowledges support and 
resources from the Center for High 
Performance Computing  at the 
University of Utah. The SDSS 
website is www.sdss4.org.

SDSS-IV is managed by the 
Astrophysical Research Consortium 
for the Participating Institutions 
of the SDSS Collaboration including 
the Brazilian Participation Group, 
the Carnegie Institution for Science, 
Carnegie Mellon University, Center for 
Astrophysics | Harvard \& 
Smithsonian, the Chilean Participation 
Group, the French Participation Group, 
Instituto de Astrof\'isica de 
Canarias, The Johns Hopkins 
University, Kavli Institute for the 
Physics and Mathematics of the 
Universe (IPMU) / University of 
Tokyo, the Korean Participation Group, 
Lawrence Berkeley National Laboratory, 
Leibniz Institut f\"ur Astrophysik 
Potsdam (AIP),  Max-Planck-Institut 
f\"ur Astronomie (MPIA Heidelberg), 
Max-Planck-Institut f\"ur 
Astrophysik (MPA Garching), 
Max-Planck-Institut f\"ur 
Extraterrestrische Physik (MPE), 
National Astronomical Observatories of 
China, New Mexico State University, 
New York University, University of 
Notre Dame, Observat\'ario 
Nacional / MCTI, The Ohio State 
University, Pennsylvania State 
University, Shanghai 
Astronomical Observatory, United 
Kingdom Participation Group, 
Universidad Nacional Aut\'onoma 
de M\'exico, University of Arizona, 
University of Colorado Boulder, 
University of Oxford, University of 
Portsmouth, University of Utah, 
University of Virginia, University 
of Washington, University of 
Wisconsin, Vanderbilt University, 
and Yale University.

\section*{Data Availability}

This work relies on publicly available data from APOGEE \citep{APOGEE_DR17} and \textit{Gaia} \citep{gaia_mission, gaia_dr3}. The catalog of abundance predictions from the \textit{Gaia} BP/RP spectra from \citet{Kane_2024} is \href{}{available on Zenodo}. The \href{https://people.smp.uq.edu.au/HolgerBaumgardt/globular/}{catalog of globular clusters} is also publicly available. The code will all be posted on GitHub following publication.



\bibliographystyle{mnras}
\bibliography{bibliography} 





\bsp	
\label{lastpage}
\end{document}